\documentclass[traditabstract]{aa} 
\usepackage{graphicx}
\usepackage{natbib} 
\usepackage[utf8]{inputenc}

\usepackage{bm} 
\usepackage{url} 
\usepackage{xcolor}
\usepackage{multirow} 
\usepackage[normalem]{ulem}

\newcommand{\emm}[1]{\ensuremath{#1}}
\newcommand{\emr}[1]{\emm{\mathrm{#1}}}
\newcommand{\unit}[1]{\emr{\,#1}} 
\newcommand{\pc}{\unit{pc}}
\newcommand{\kpc}{\unit{kpc}}

\newcommand{\mum}{\unit{\mu m}}

\newcommand{\pscm}{\unit{cm^{-2}}} 

\newcommand{\kms}{\unit{km\,s^{-1}}}
\newcommand{\Kkms}{\unit{K\,km\,s^{-1}}} 
\newcommand{\K}{\unit{K}}

\newcommand{\GHz}{\unit{GHz}}

\newcommand{\Msun}{\unit{M_{\sun}}}

\newcommand{\Msunpsqpc}{\unit{M_{\sun}/pc^2}}

\newcommand{\Xco}{\emm{X_\emr{CO}}} 
\newcommand{\XHi}{\emm{X_\emr{\Hi}}}

\newcommand{\pscmpKkms}{\unit{cm^{-2}/(K\,km\,s^{-1})}}
\newcommand{\chem}[1]{\ensuremath{\mathrm{#1}}}
 
\newcommand{\Ht}{\chem{H_{2}}}
\newcommand{\Hy}{\chem{H}} 
\newcommand{\Hi}{\ion{H}{i}}
\newcommand{\He}{\chem{He}} 

\newcommand{\twCO}{\chem{^{12}CO}}

\newcommand{\Jtwo}{(2--1)}

\def\C3H2{c-\chem{C_{3}H_{2}}}

\newcommand{\sciexp}[2]{\emm{#1\times10^{#2}}}

\newcommand{\N}[1]{\emm{N(#1)}}
\newcommand{\I}[1]{\emm{I_{#1}}} 
\newcommand{\NH}{\N{\Hy}}
\newcommand{\NHt}{\N{\Ht}} 
\newcommand{\NHtot}{\N{H}} 

\newcommand{\NHi}{\N{\Hi}}

\newcommand{\aco}{\emm{\alpha_{\emr{CO}}}}
\newcommand{\xco}{\emm{X_{\emr{CO}}}}
\newcommand{\ahi}{\emm{\alpha_{\ion{H}{i}}}}
\newcommand{\gdr}{\emm{{GDR}}} 
\newcommand{\kdark}{\emm{K_{\emr{dark}}}}
\newcommand{\kdarkp}{\emm{K^{\prime}_{\emr{dark}}}}
\newcommand{\ico}{\I{\emr{CO}}} 
\newcommand{\ihi}{\emm{I_{\ion{H}{i}}}}
\newcommand{\icoi}{\emm{I_{\emr{CO},i}}}
\newcommand{\ihii}{\emm{I_{\ion{H}{i},i}}}
\newcommand{\Shi}{\emm{\Sigma_{\ion{H}{i}}}}
\newcommand{\Sht}{\emm{\Sigma_{\Ht}}}
\newcommand{\Sdust}{\emm{\Sigma_{dust}}}
\newcommand{\Sdusti}{\emm{\Sigma_{dust,i}}}
\newcommand{\Sgas}{\emm{\Sigma_{gas}}}

\newcommand{\ratioo}{{\N{\Ht}/\I{\emr{CO}}}}
\newcommand{\Xunit}{\pscm/\Kkms}
\newcommand{\aunits}{\unit{M_{\sun}/pc^2/\Kkms}}
\newcommand{\sigmadust}{\emm{\sigma_{dust}}}
\newcommand{\Zgas}{\emr{Z_{gas}}} 
\usepackage[]{hyperref}

\newcommand{\FigNHtICOnocutsub}{ 
\begin{figure}
	\flushleft
	\includegraphics[width=.9\hsize{}]{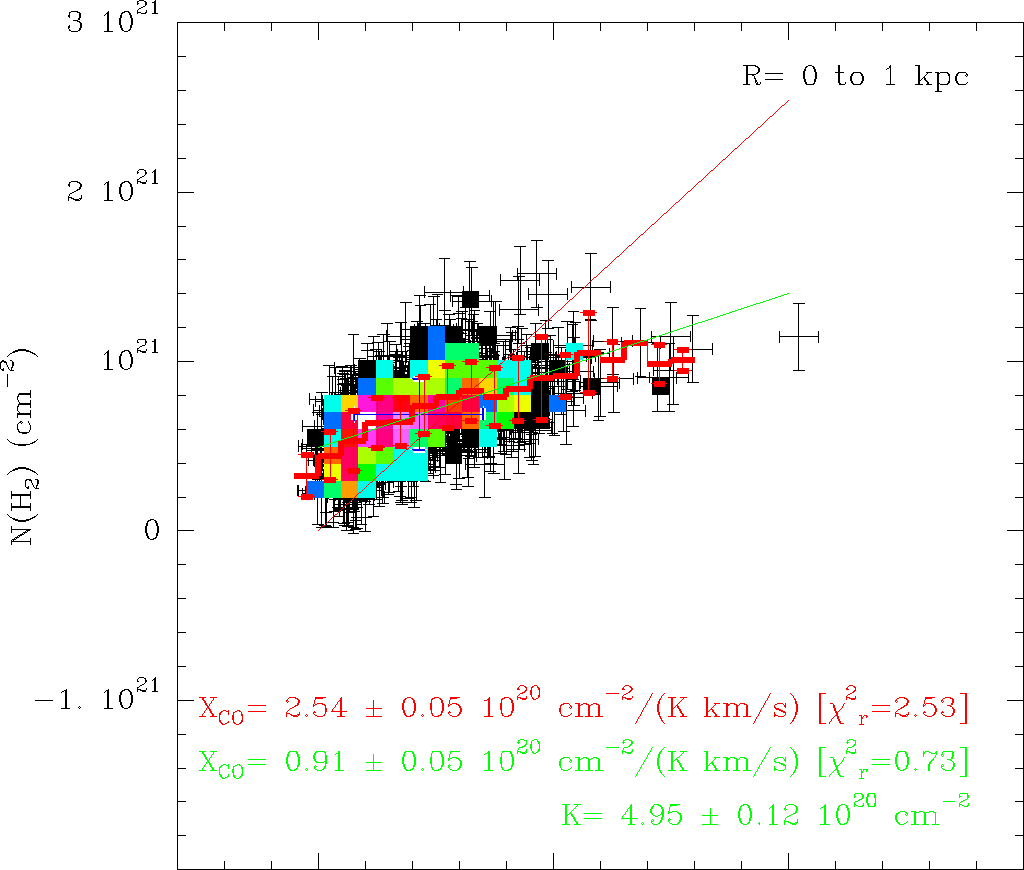}
	\includegraphics[width=.9\hsize{}]{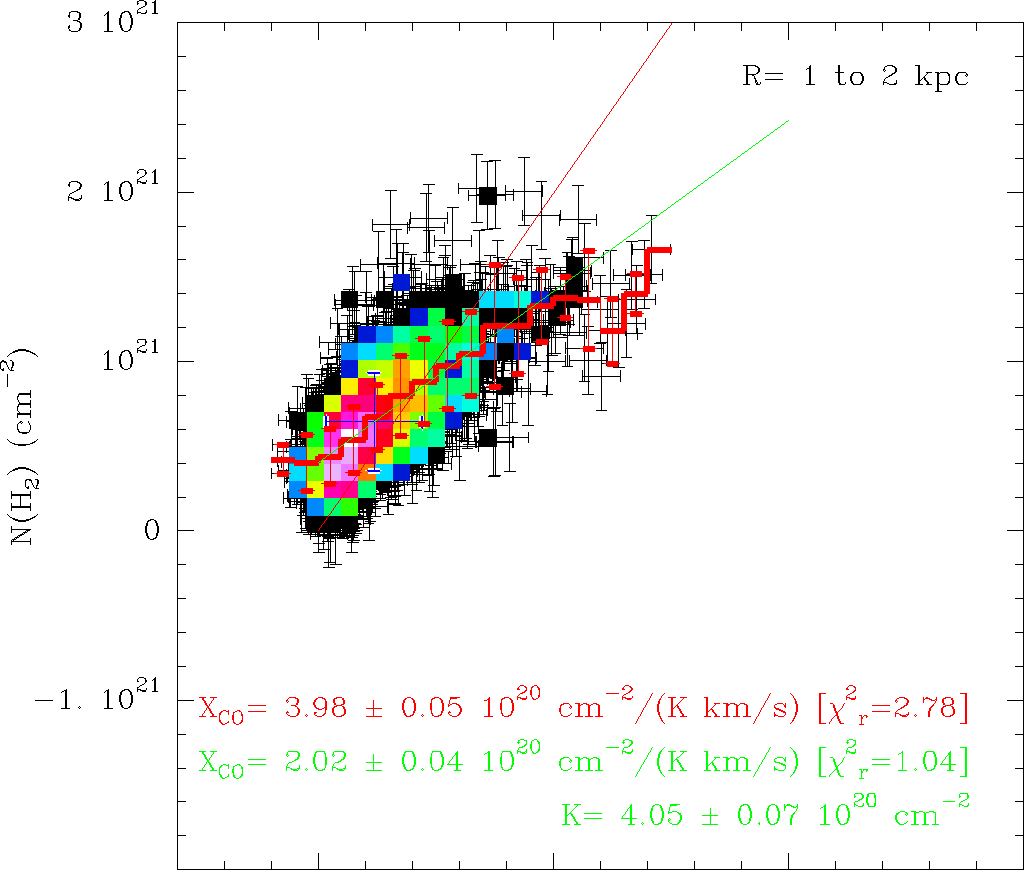}
	\includegraphics[width=.915\hsize{}]{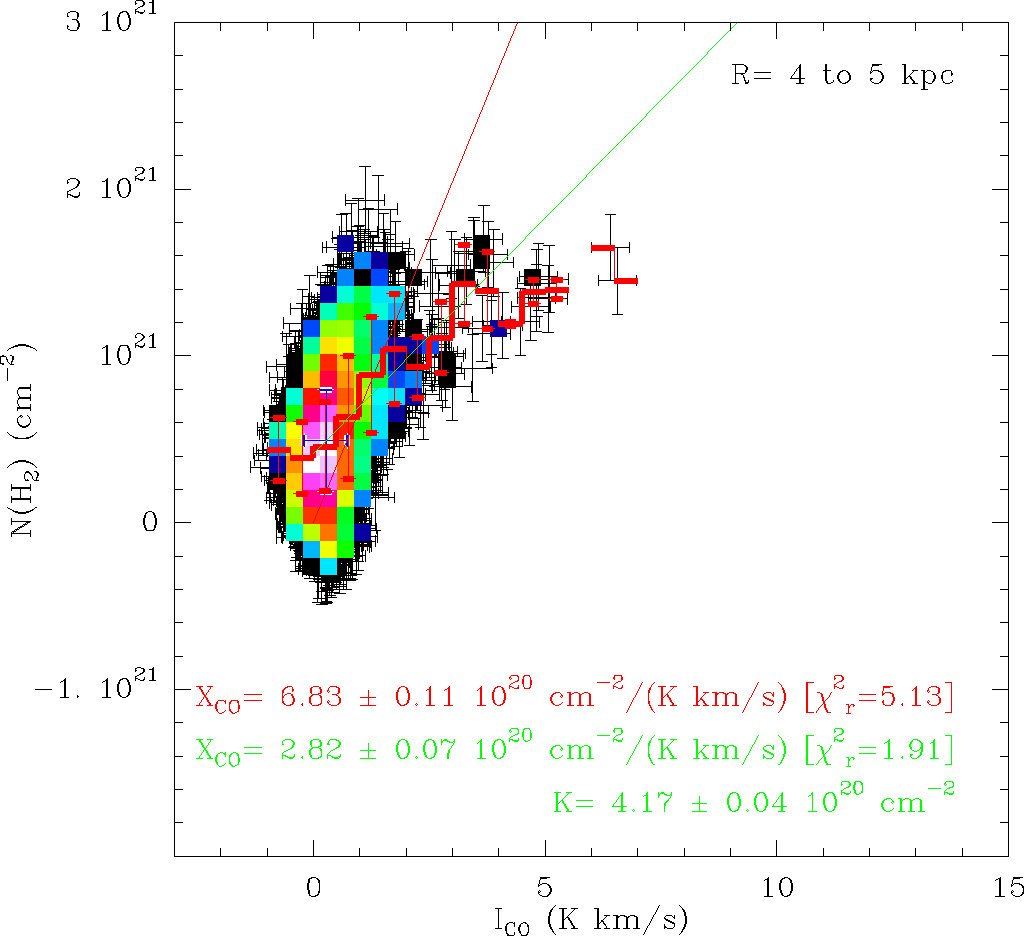}
	\caption{Fit of dust-derived \NHt\ as a function of \ico\ for data in
	radial intervals between 0 and 1 \kpc\ ($top$), $1-2$\kpc\ ($middle$),
	and $4-5$\kpc\ ($bottom$). No cut in intensity has been applied. The
	color scale indicates the density of points and the thick red histogram
	shows the \NHt\ data averaged in bins of 0.5 \unit{K km/s}. The thin
	green line shows an affine fit between \NHt\ and \ico; the corresponding
	fit results are printed in green. The thin red line is a linear fit
	without an offset; the corresponding fit results are printed in red.
	Blue cross:
	Average value of the plotted data.} 
	\label{fig.nht_ico_nocut_sub}
\end{figure}
}

\newcommand{\FigtwoDmeansub}{ 
\begin{figure}
	\centering
	\includegraphics[width=0.9\hsize{}]{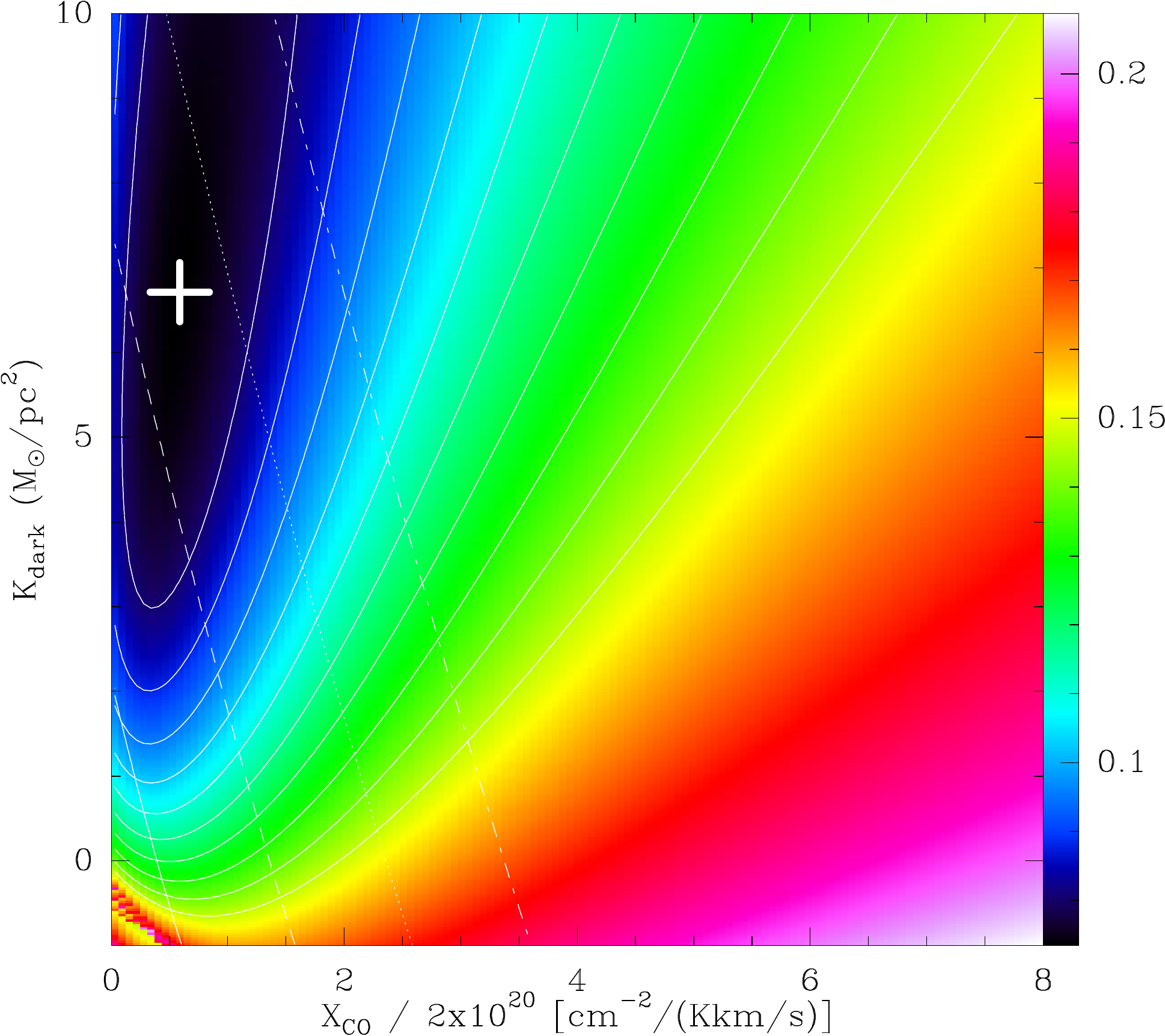}
	\includegraphics[width=0.9\hsize{}]{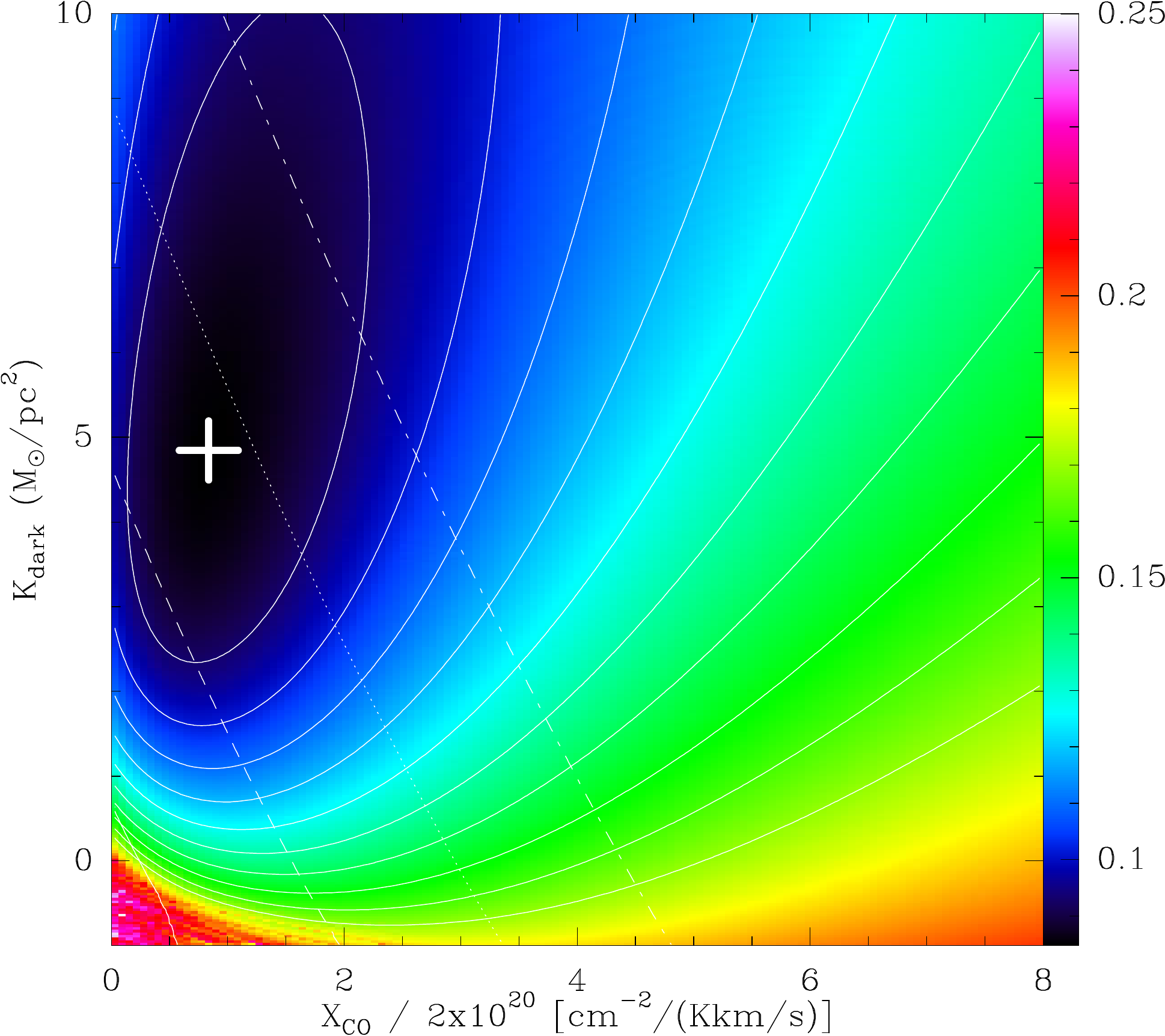}
	\includegraphics[width=0.9\hsize{}]{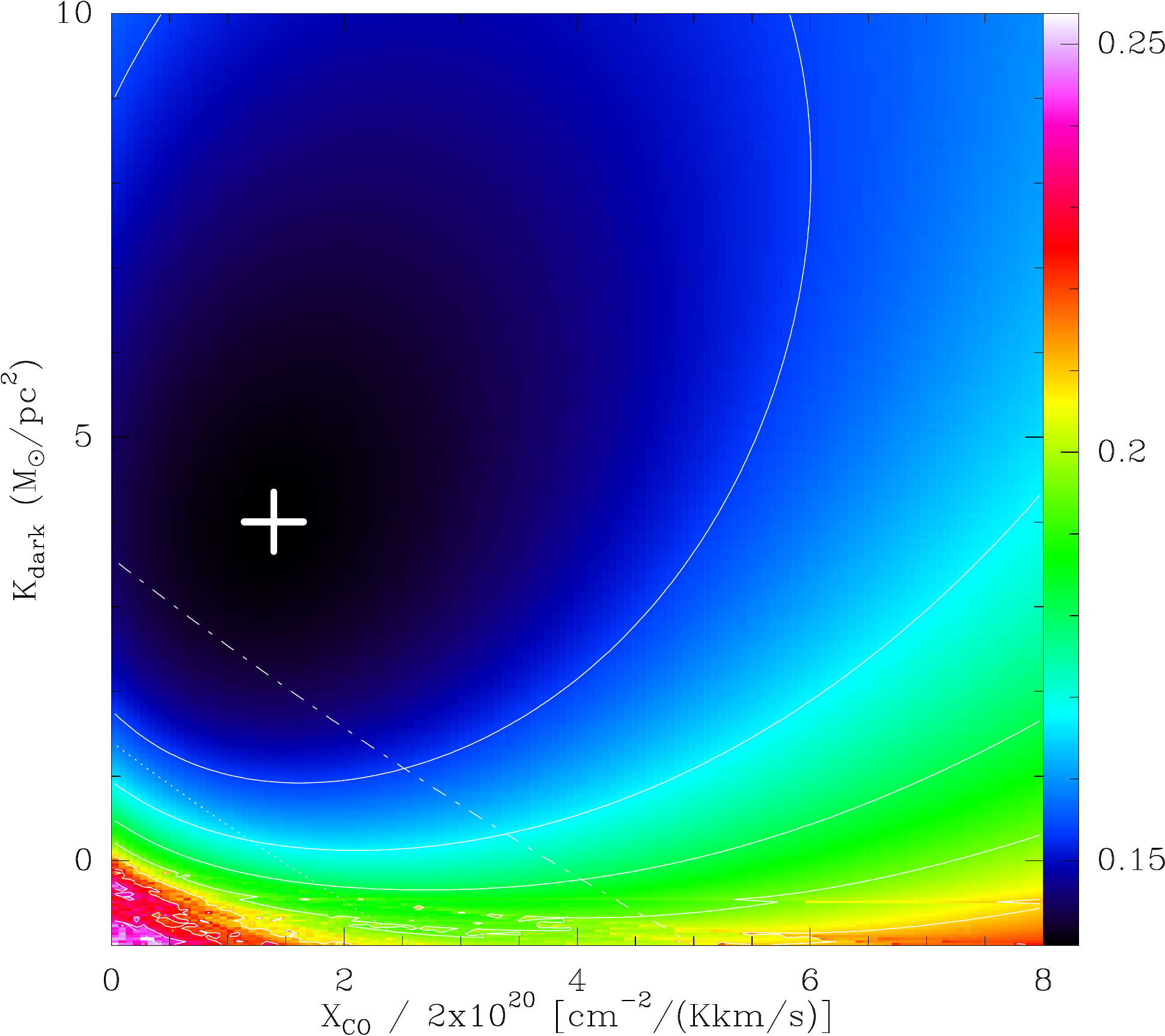}
	\caption{Scatter in $\log(\gdr)$ as a function of \xco\ and \kdark.
	The color scale and solid white contours indicate the amplitude of the
	scatter in $\log(\gdr)$ as measured by the standard deviation for
	varying \xco\ and \kdark\ offsets. radii between 0 and 1 kpc ($top$),
	$1-2$kpc ($middle$), and $4-5$kpc ($bottom$).The white cross corresponds
	to the minimum scatter (i.e., best fit). The contours correspond to
	constant scatter values and give an indication of the uncertainties and
	degeneracies. The white lines correspond to constant \gdr\ values of 100
	(solid), 150 (dashed), 200 (dotted), 250 (dash-dotted).}
	\label{fig.scatter_mean_sub} 
\end{figure}
}

\newcommand{\FigSimu}{ 
\begin{figure*}
	\begin{centering}
		\includegraphics[width=0.495\hsize{}]{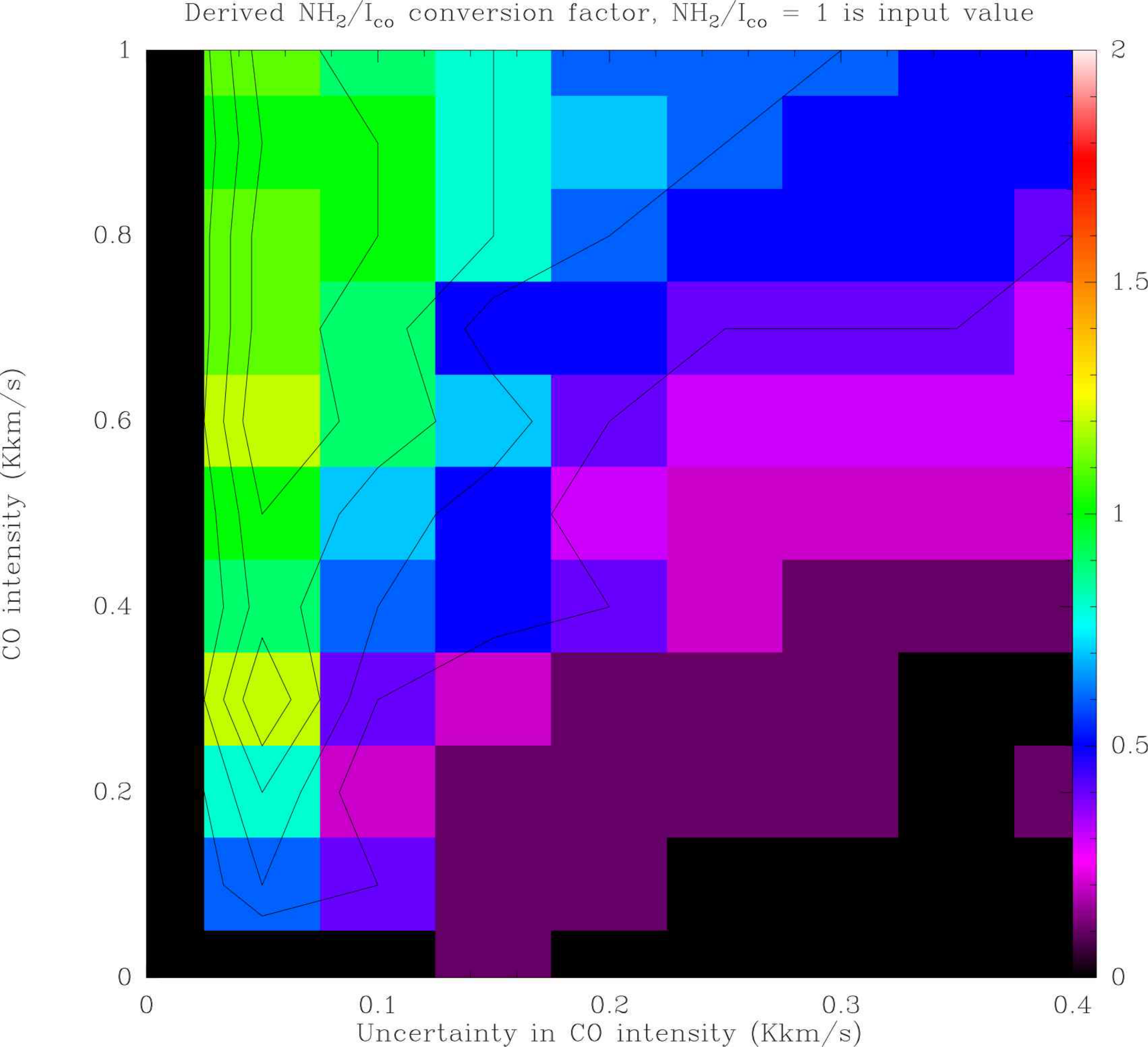}
		\includegraphics[width=0.495\hsize{}]{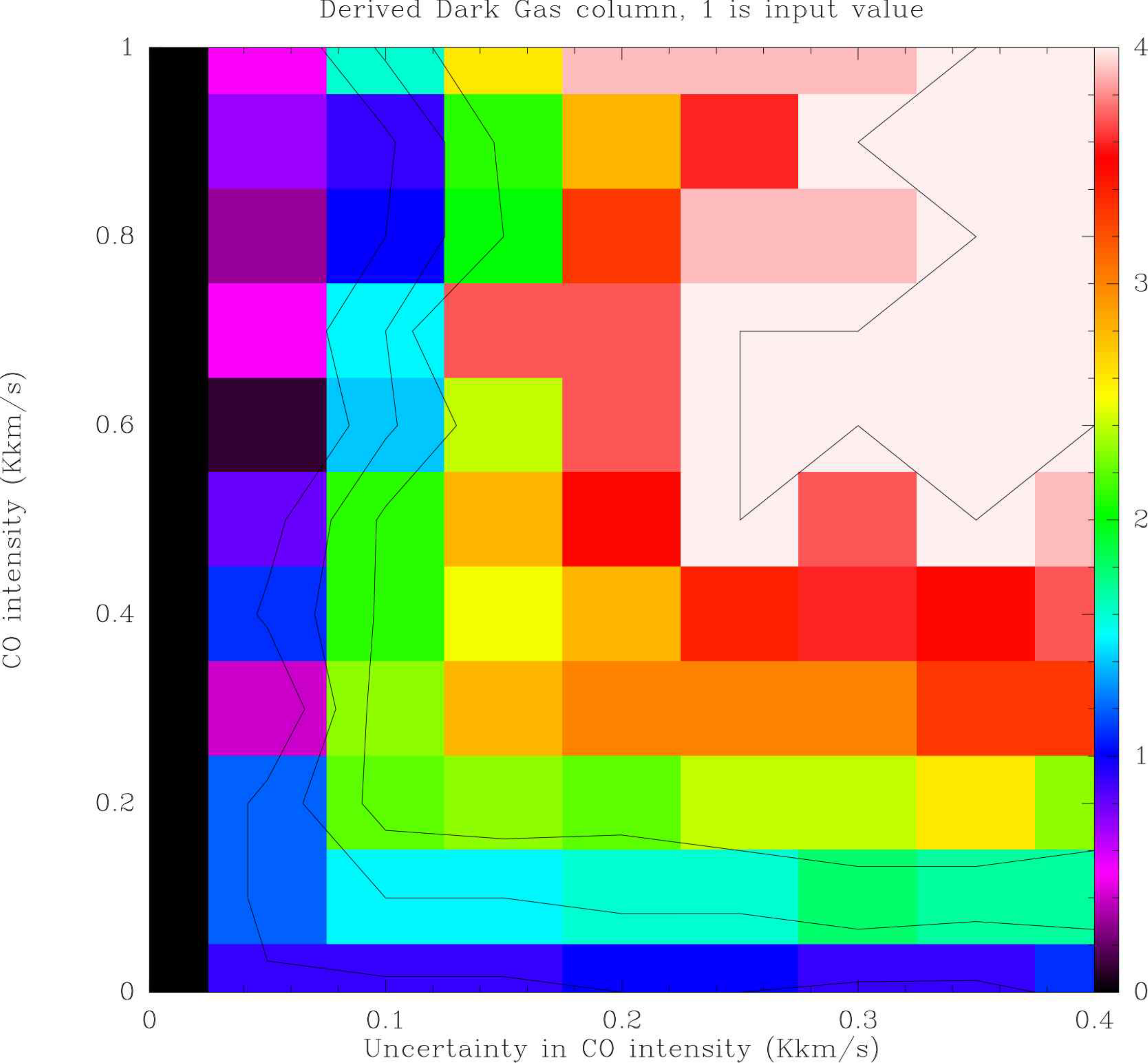}
		\includegraphics[width=0.495\hsize{}]{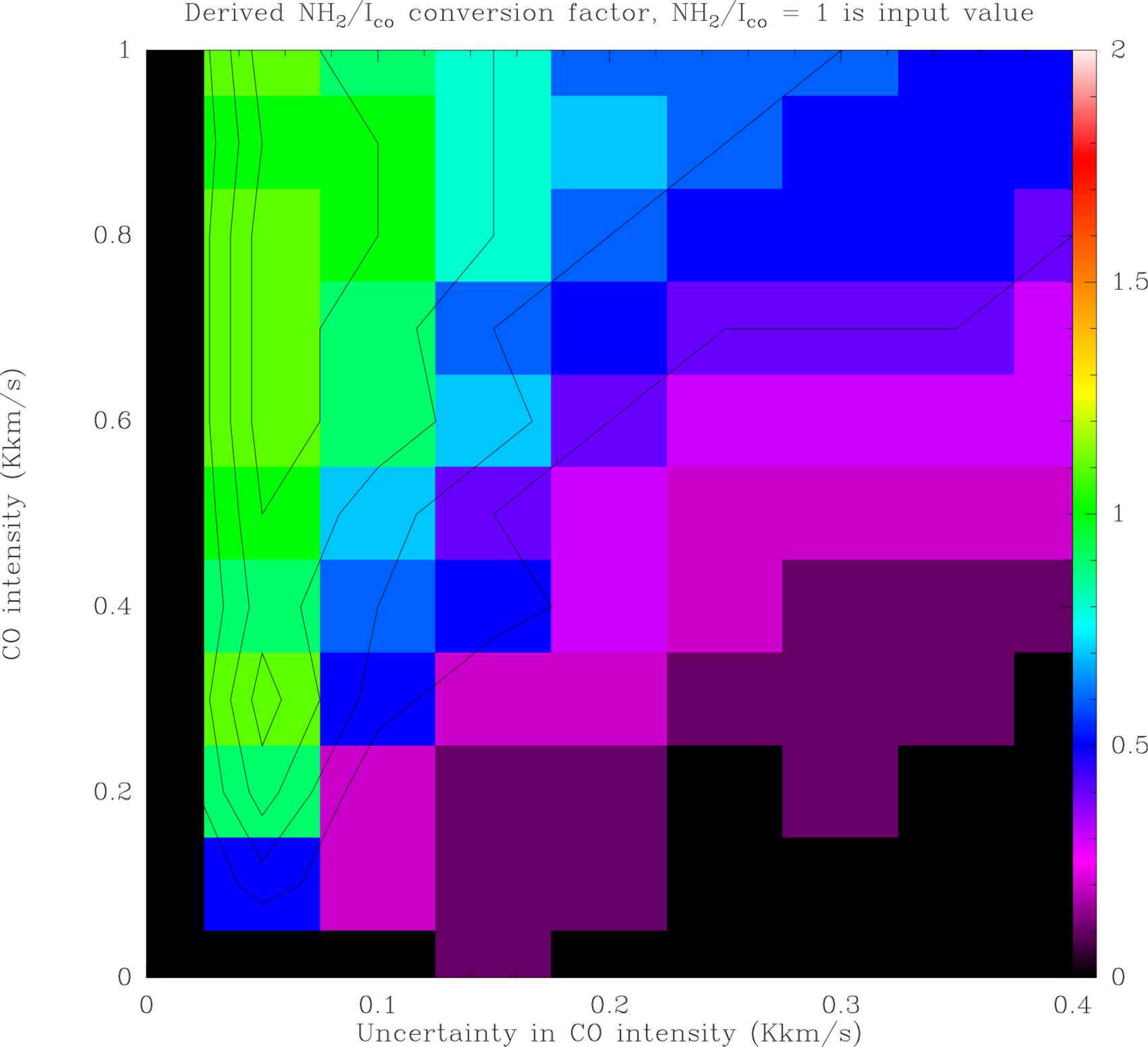}
		\includegraphics[width=0.495\hsize{}]{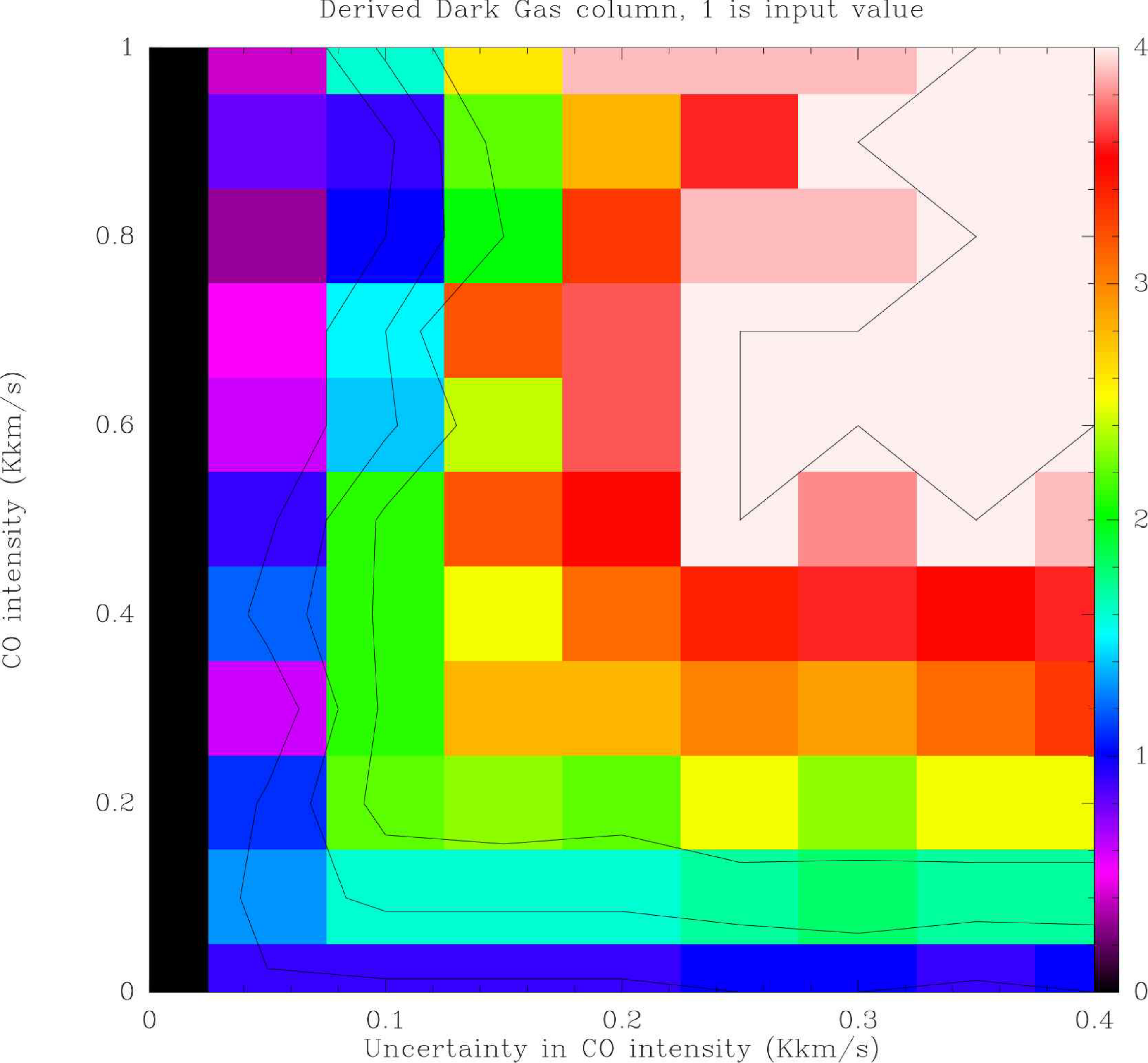}
		\includegraphics[width=0.495\hsize{}]{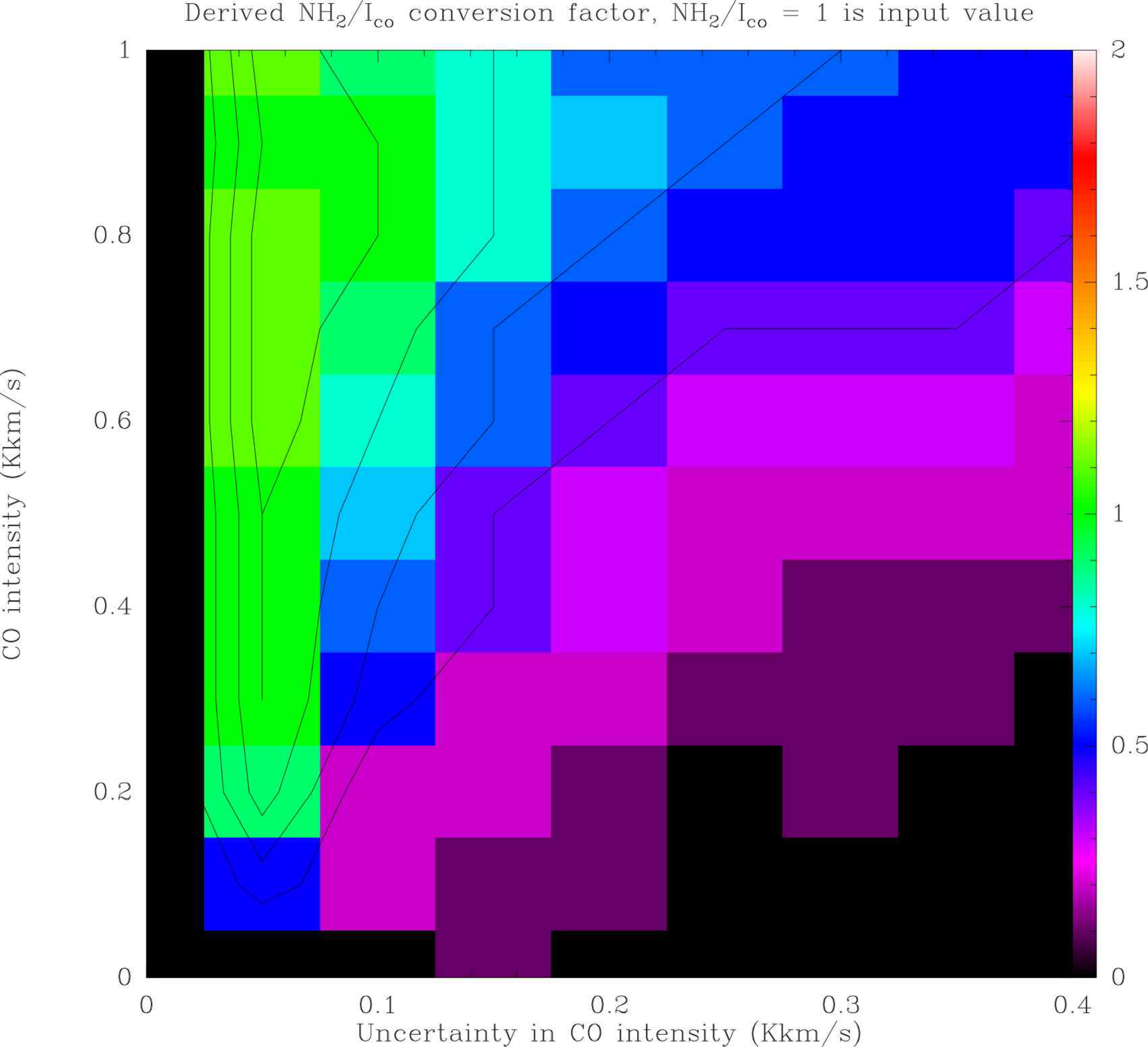}
		\includegraphics[width=0.495\hsize{}]{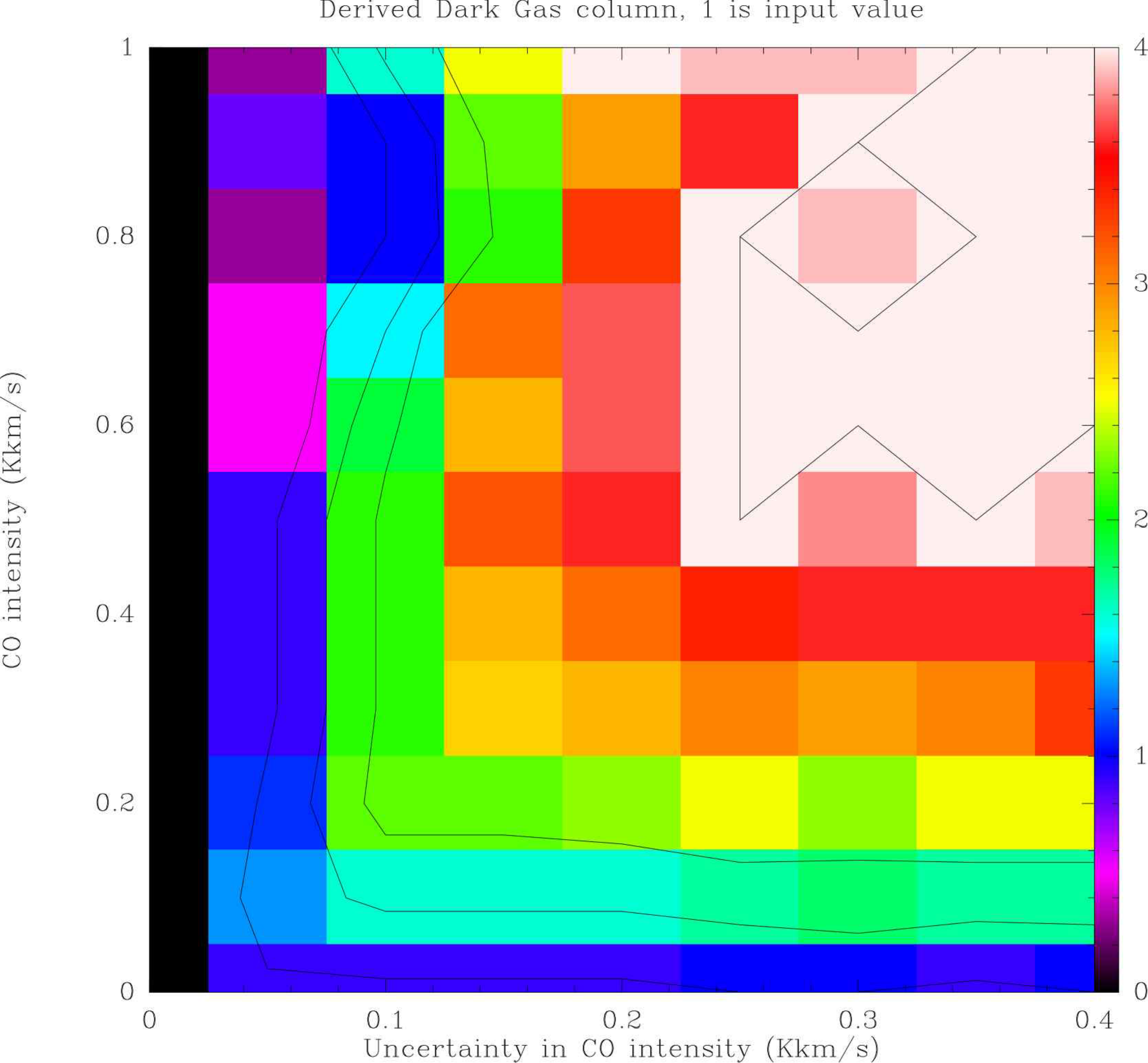}
	\end{centering}
	\caption{Optimal retrieved values of $\xco$ (left
	column) and K (right column) as a function of the CO intensity (before
	adding noise) and the noise level of the CO observations. Top figures:
	fiducial model.
	Middle row: fiducial except \Hi\ column density reduced to $4 \times
	10^{20}$cm$^{-2}$. Bottom: fiducial except dust uncertainties reduced to
	10\%.} 
	\label{fig.LSimu} 
\end{figure*}
}

\newcommand{\FigBayesTest}{ 
\begin{figure}
	\centering
	\includegraphics[width=\hsize{}]{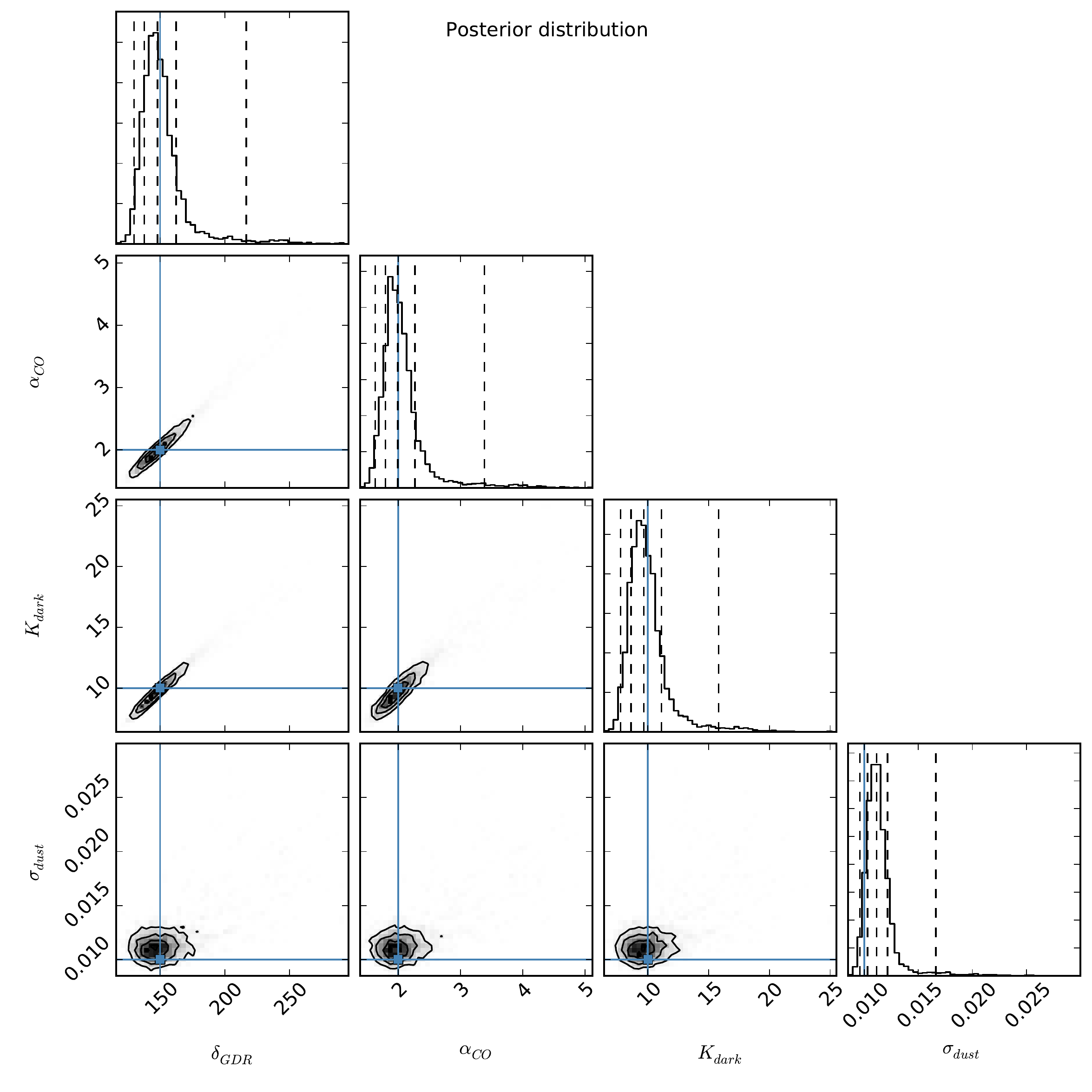} 
	\caption{Test of
	the Bayesian method. The input values for the simulation are shown as
	blue lines and these correspond rather well to the peaks of probability
	distributions determined by the method. The dashed lines indicate the
	median and $\pm 1\sigma$ and $\pm 2\sigma$ intervals. }
	\label{fig.BayeTest} 
\end{figure}
}

\newcommand{\FigLSCenter}{ 
\begin{figure}
	\centering
	\includegraphics[width=\hsize{}]{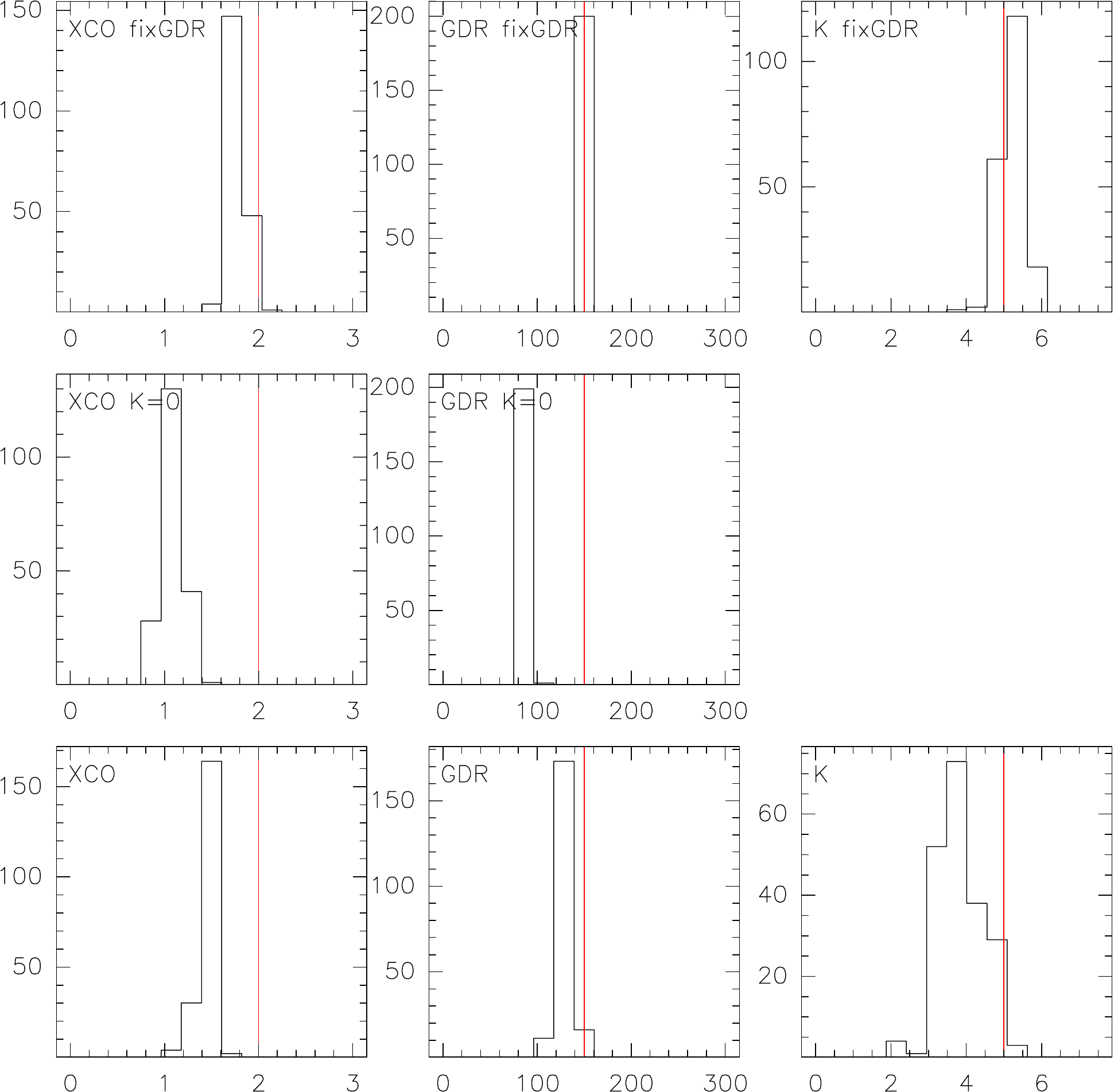}
	
	\caption{Histogram of recovered values for the generative model
	including noise in all three observables \xco, \gdr, and \kdark. Bottom
	row: recovering the 3 parameters, Middle row: recovering only \aco and
	\gdr\ even though \kdark is present in the data. Top row: same as bottom
	row but imposing the correct value of \gdr.
	This figure is for the central kpc of M33. Input values are in red.}
	
	\label{fig.LSCenter} 
\end{figure}
} 
\newcommand{\FigLSOuterNoCut}{
\begin{figure}
	\centering
	\includegraphics[width=\hsize{}]{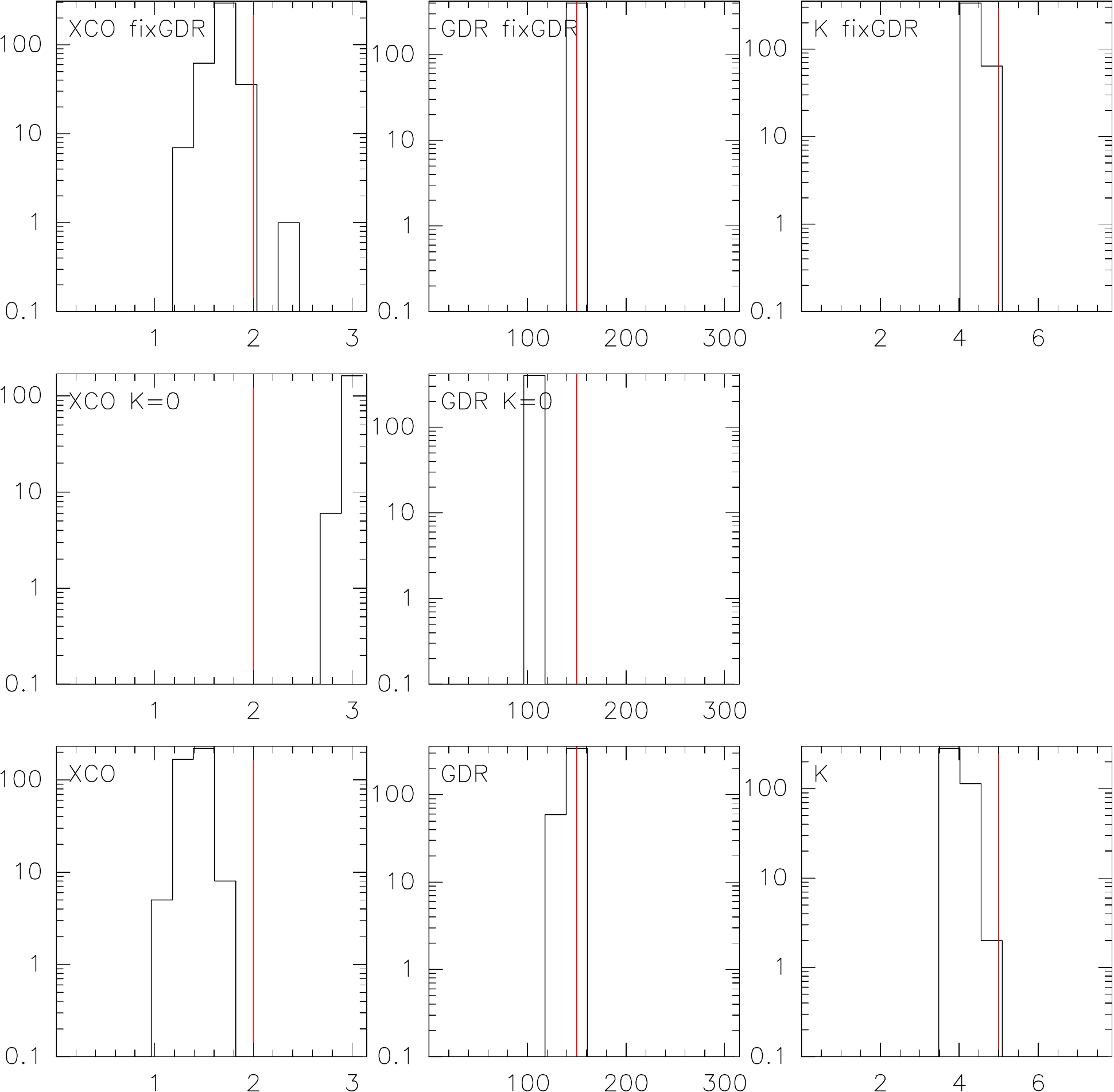} 
	\caption{Same as
	Fig.~\ref{fig.LSCenter} but for $4\kpc<R<5\kpc$}
	\label{fig.LSOuterNoCut} 
\end{figure}
} 
\newcommand{\FigLSOuterCut}{
\begin{figure}
	\centering
	\includegraphics[width=\hsize{}]{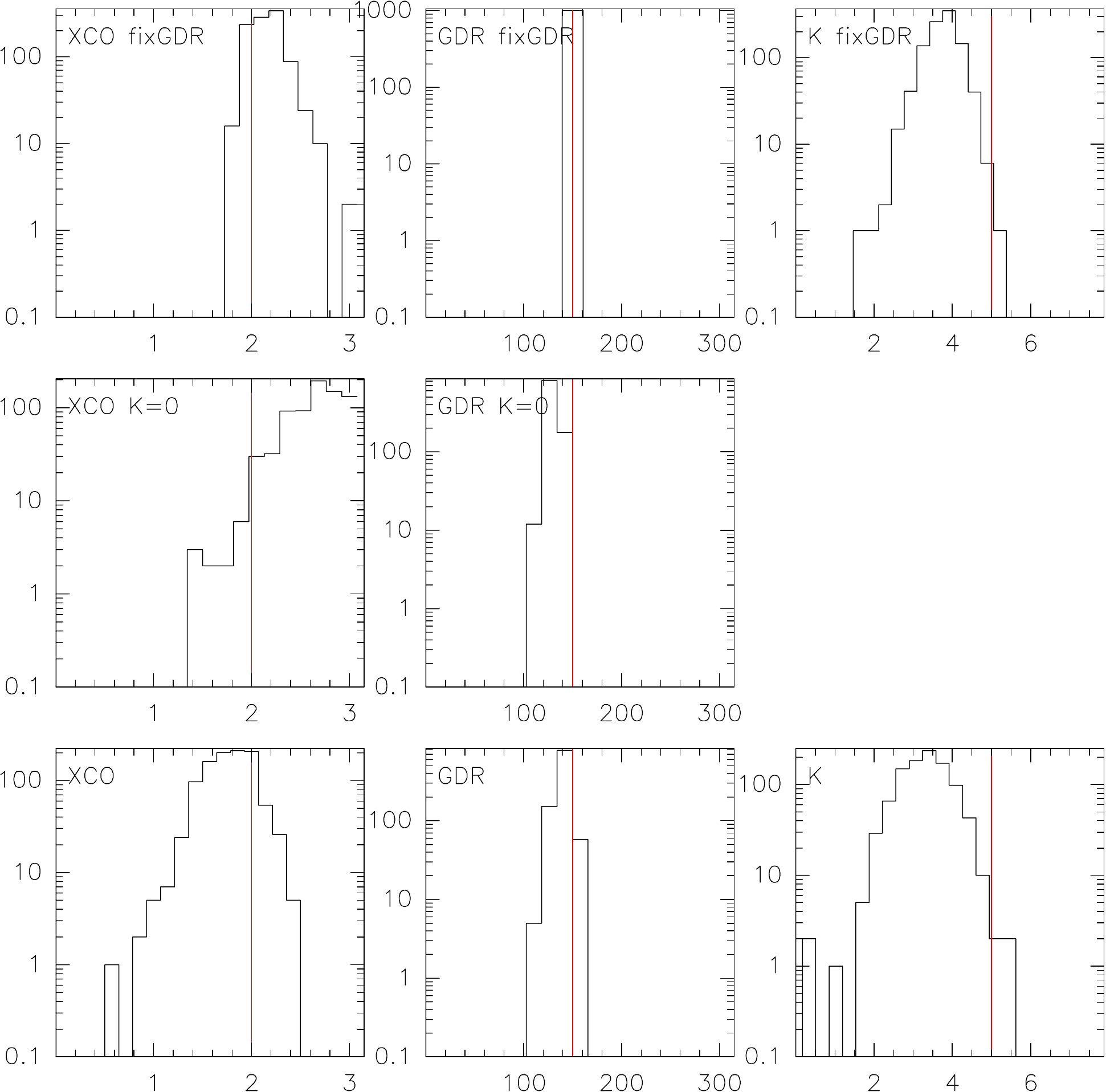}
	\caption{Same as Fig.~\ref{fig.LSOuterNoCut} but only considering pixels
	where $\ico > 2\sigma$. A similar cut for the central radii would show
	little effect as the CO signal there is strong.} 
	\label{fig.LSOuterCut}
\end{figure}
}

\newcommand{\FigDust}{ 
\begin{figure*}
	\centering
	\includegraphics[width=0.49\hsize{}]{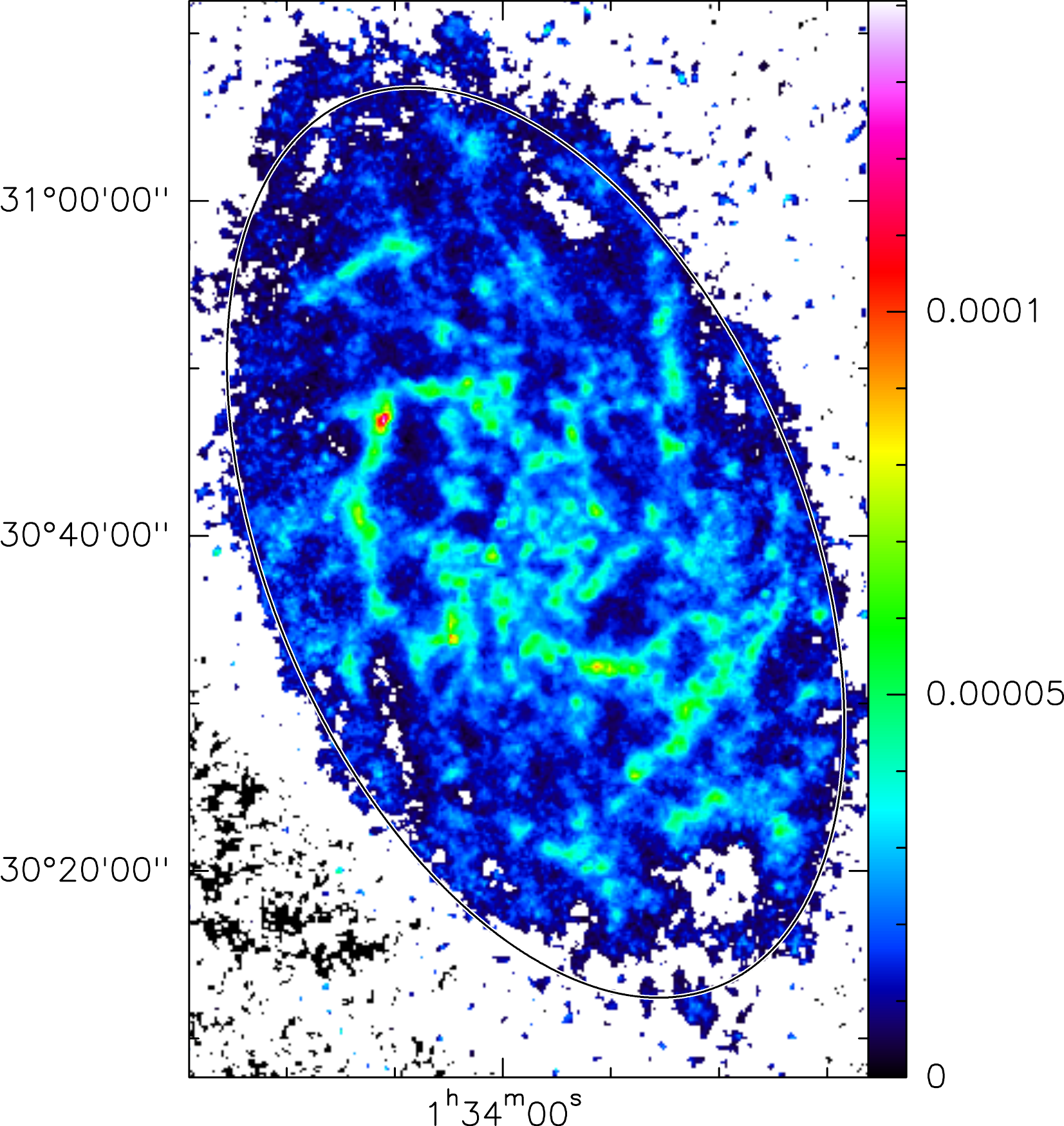}
	\includegraphics[width=0.49\hsize{}]{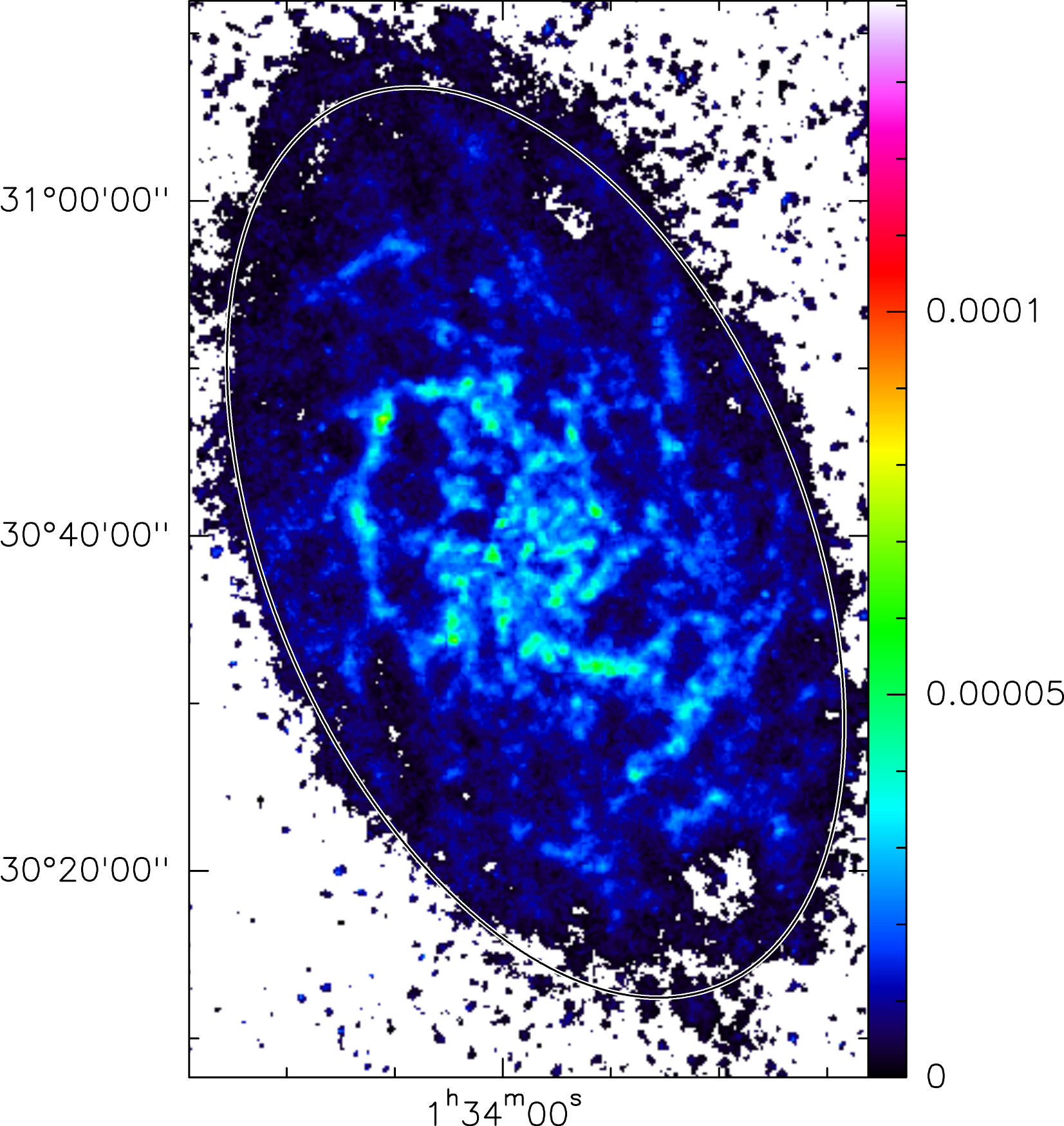} 
	\caption{Dust
	surface density [\unit{g/,cm^{-2}}] maps of M33 at 25" resolution:
	(left) for a constant $\beta=2$, (right) radially variable $\beta$
	(2-1.3) as derived in \citet{Tabatabaei.2014}. The ellipsed correspond
	to a galactocentric radius of 7\kpc.} 
	\label{fig.dust} 
\end{figure*}
}

\newcommand{\FigRadial}{ 
\begin{figure}
	\centering
	\includegraphics[width=\hsize{}]{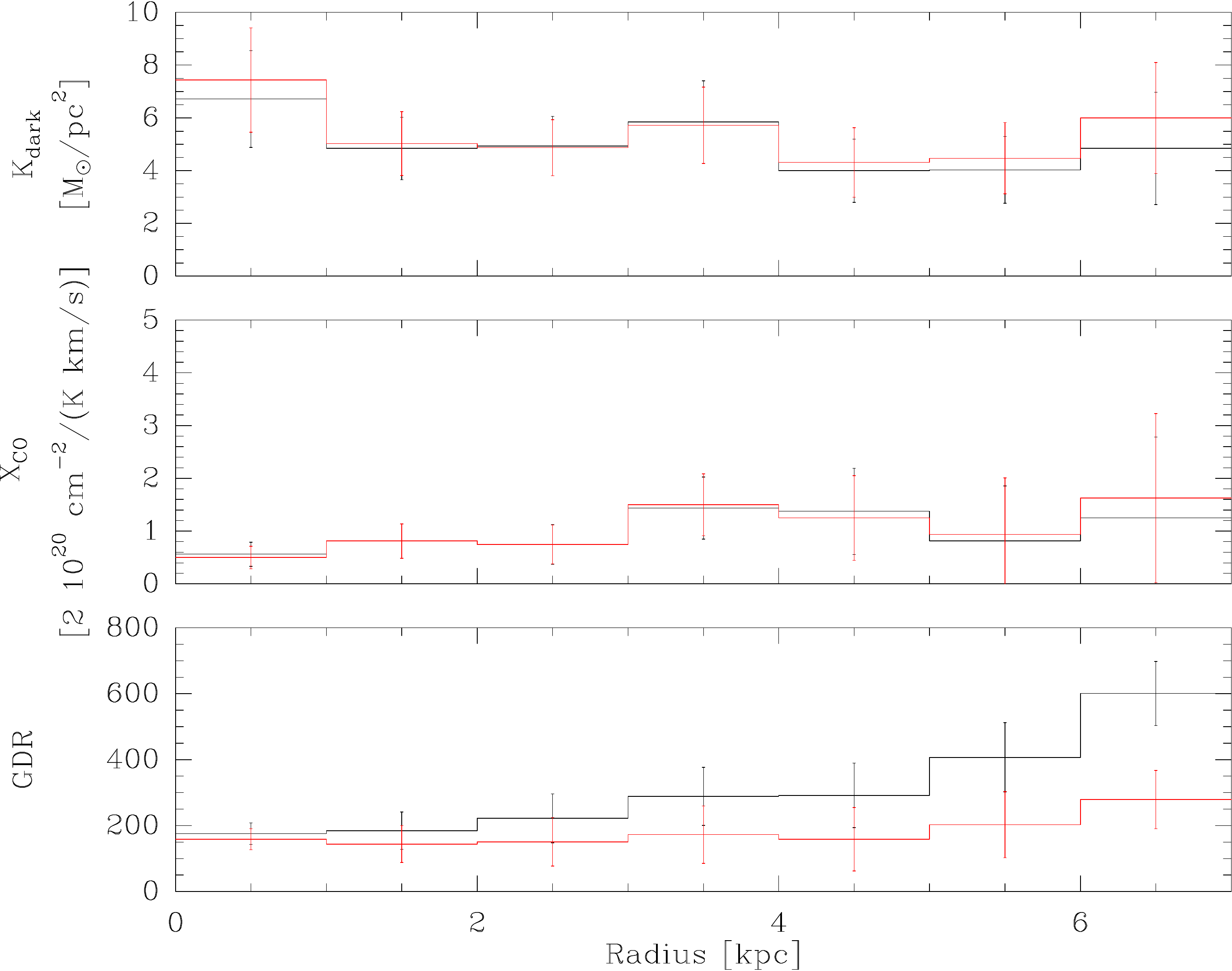} 
	\caption{Average
	values for \kdark\ , $\xco$, and \gdr\ derived for 1 kpc radial bins
	using the Leroy-Sandstrom method. The black histogram shows results
	derived with the variable beta \citet{Tabatabaei.2014} prescription and
	the red used the $\beta = 2$ to determine dust temperatures.}
	\label{fig.radial} 
\end{figure}
}

\newcommand{\FigRadialBayesCompare}{ 
\begin{figure}
	\centering
	\includegraphics[width=\hsize{}] {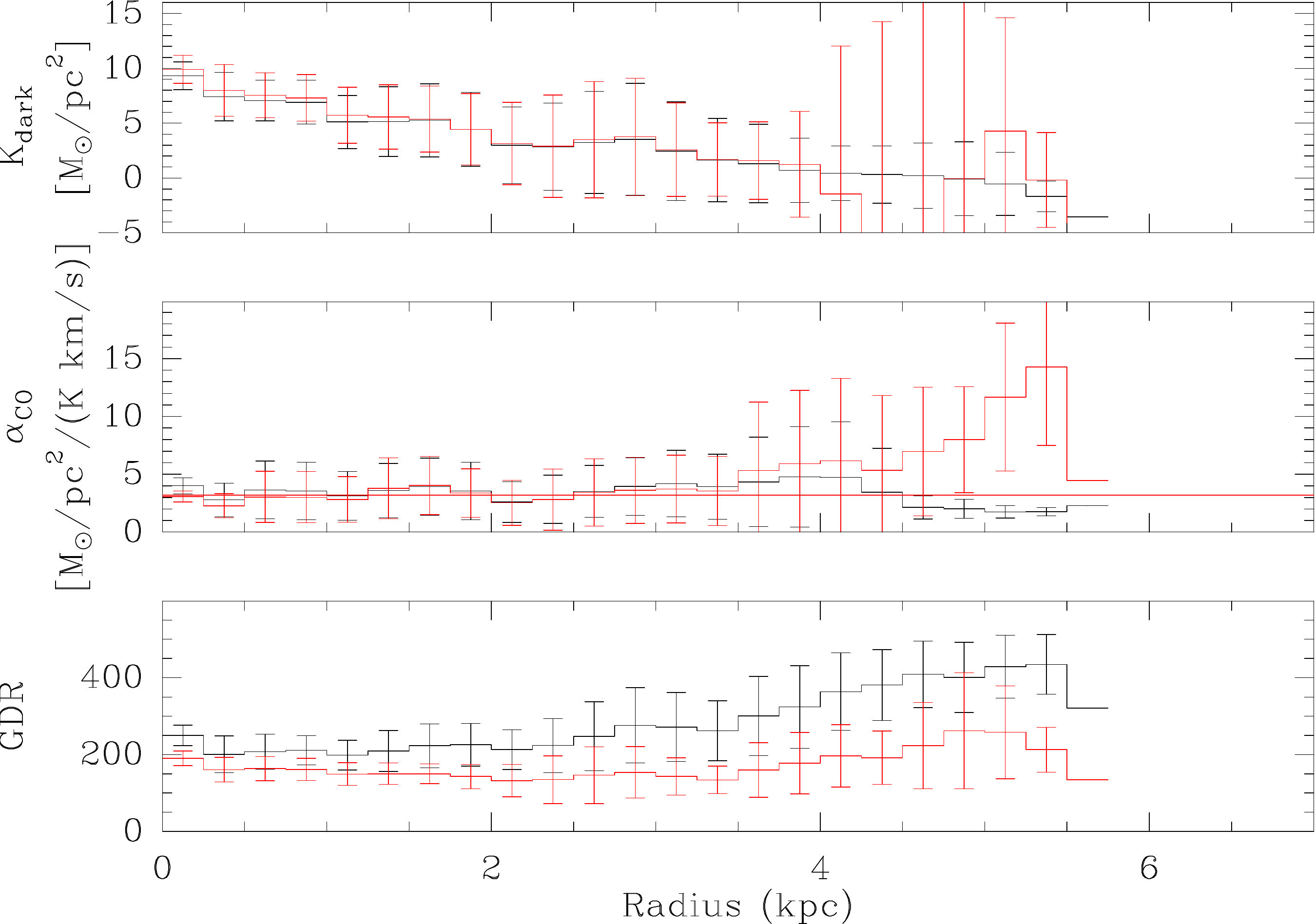} 
	\caption{Average
	values for \kdark\ , $\xco$, and \gdr\ derived for 0.5 kpc radial bins
	using the Bayesian method. The black histogram shows results derived
	with the variable beta of \citet{Tabatabaei.2014} and the red uses the
	standard $\beta = 2$ to determine dust temperatures.}
	\label{fig.radial_bayes_compare} 
\end{figure}
}

\newcommand{\FigLSnoise}{ 
\begin{figure}
	\centering
	\includegraphics[width=\hsize{}]{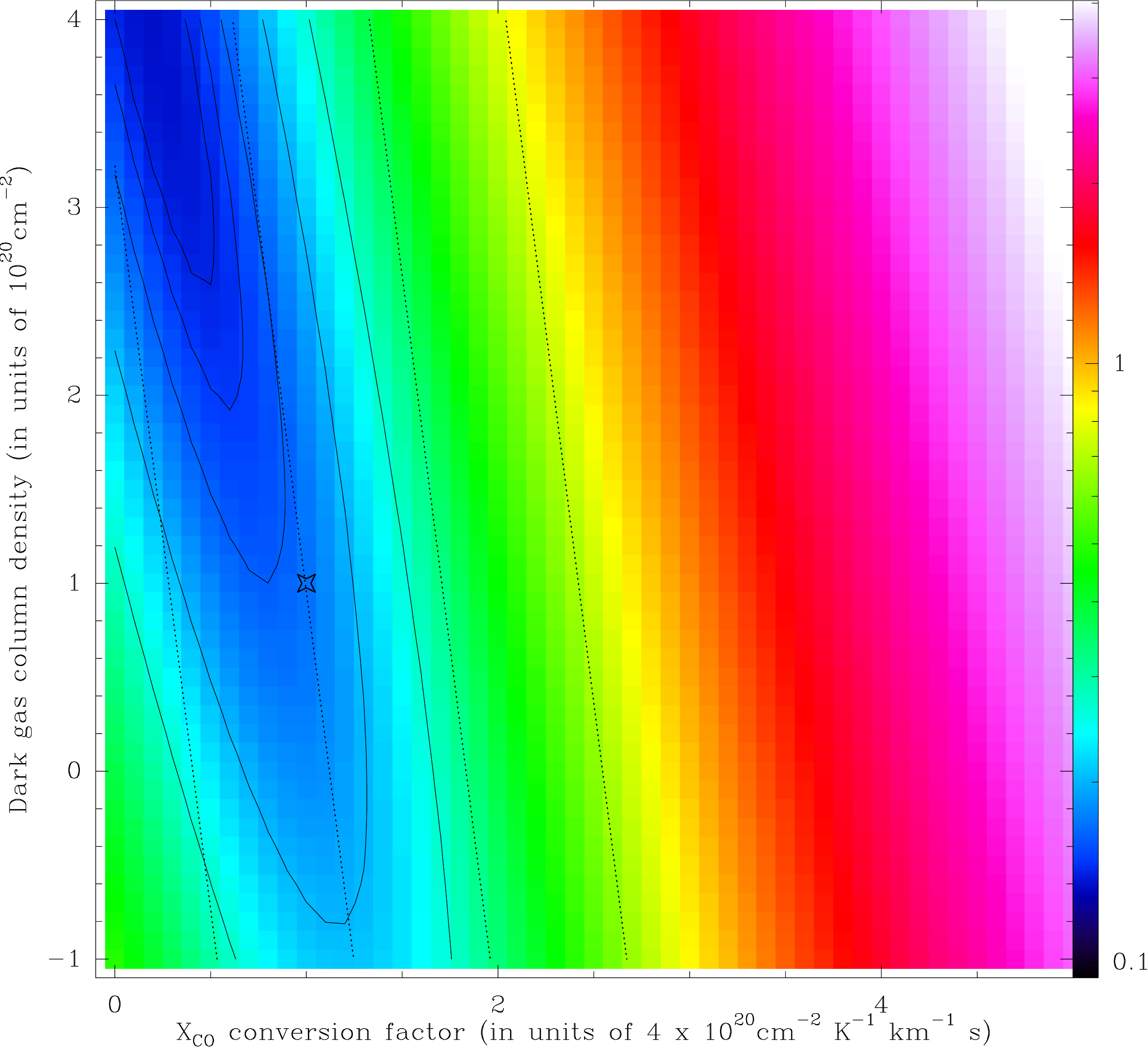}
	\caption{Quality of fit for model with $\ico = 1 \pm 0.25$ K\kms, \NHi$
	= 8 \pm 1 \times 10^{20} \pscm$, and $\kdark = 1 \pm 0.25 \times 10^{20}
	\pscm$, assuming that the uncertainty in the dust surface density is
	25\%. Dotted lines represent, from left to right, constant \gdr\ values
	of 100, 150, 200, 250. The star at $\xco = 4 \times 10^{20} \pscmpKkms$
	is the input value but the best fit is far from that.}
	\label{fig.LSnoise} 
\end{figure}
}

\newcommand{\FigCorrelNew}{ 
\begin{figure}
	\centering
	\includegraphics[width=\hsize{}]{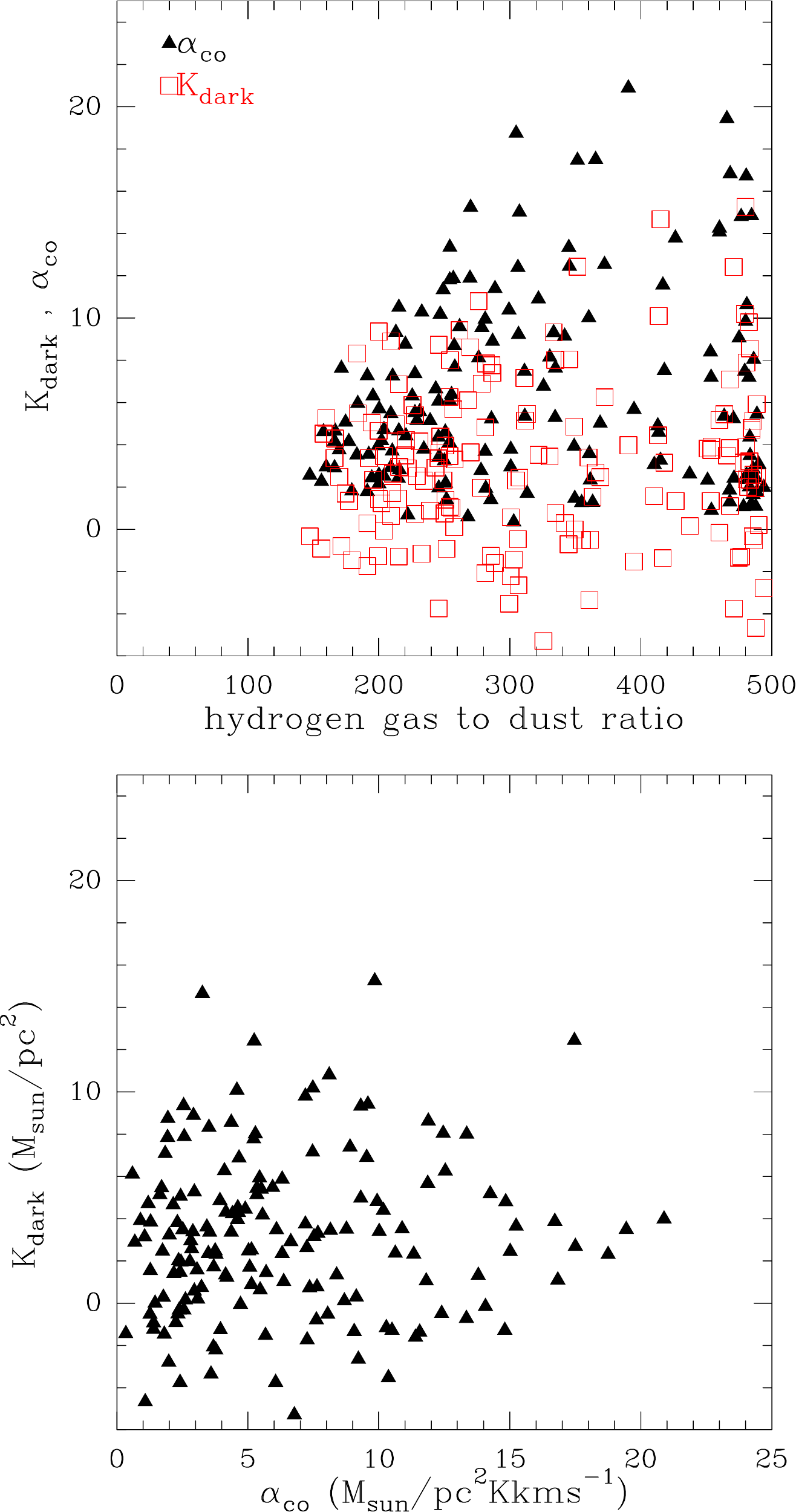} 
	\caption{Search for
	degeneracies between the \gdr, $\xco$, and $\kdark$ in the Bayesian
	approach. Top panel shows $\xco$ ($\aco$) and $\kdark$ as a function of
	\gdr. Bottom panel shows the link (or absence) between $\xco$
	($\alpha_{co}$) and $\kdark$. Each point represents a pixel in the maps
	shown in Fig.~\ref{fig.map_betavar_3sigcut_GDRprior}.}
	\label{fig.Correlnew} 
\end{figure}
}

\newcommand{\FigXK}{ 
\begin{figure}
	\centering
	\includegraphics[width=\hsize{}]{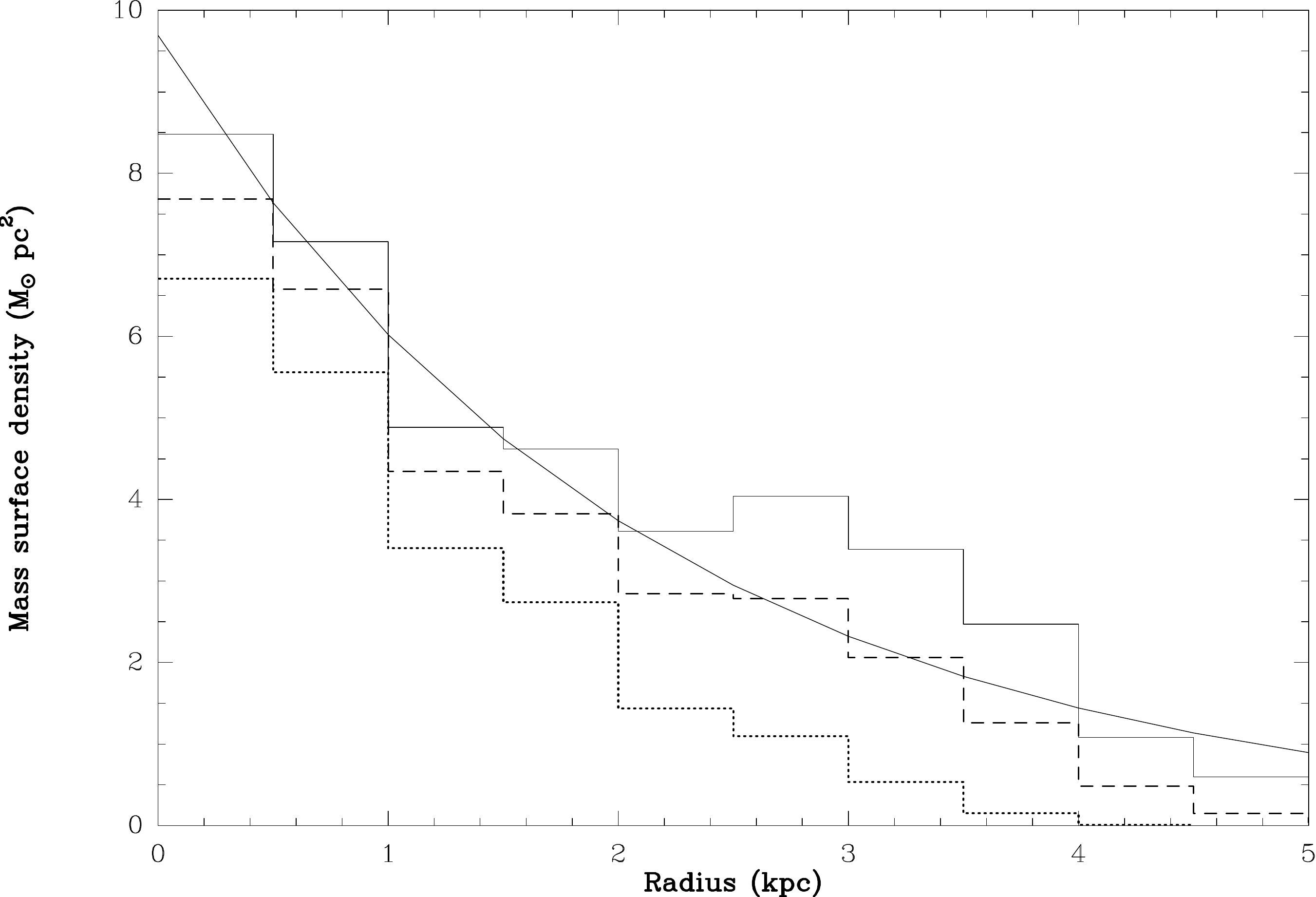} 
	\caption{H$_2$ surface
	density derived from CO and $\kdark$ derived from the Bayesian analysis.
	The continuous curve shows $\Sigma_{H_2}$ based on Fig. 10 of
	\citet{Druard.2014} corrected to a $\xco$ factor of 1.1 Galactic and
	uncorrected for inclination and helium content. The histograms show the
	CO dark gas surface density. The solid histogram shows $\kdark$ as
	derived assuming that all positions have the same dark column as the
	positions where CO is detected above the threshold. The dashed and
	dotted histograms represent $\kdark$ assuming that the dark column only
	is present where CO is detected above the $0\sigma$ and $3\sigma$
	thresholds respectively.} 
	\label{fig.XK} 
\end{figure}
}

\newcommand{\FigMapBetavarSigcutthreeGRDlimit}{ 
\begin{figure}
	\centering 
	\includegraphics[width=\hsize{}]
	{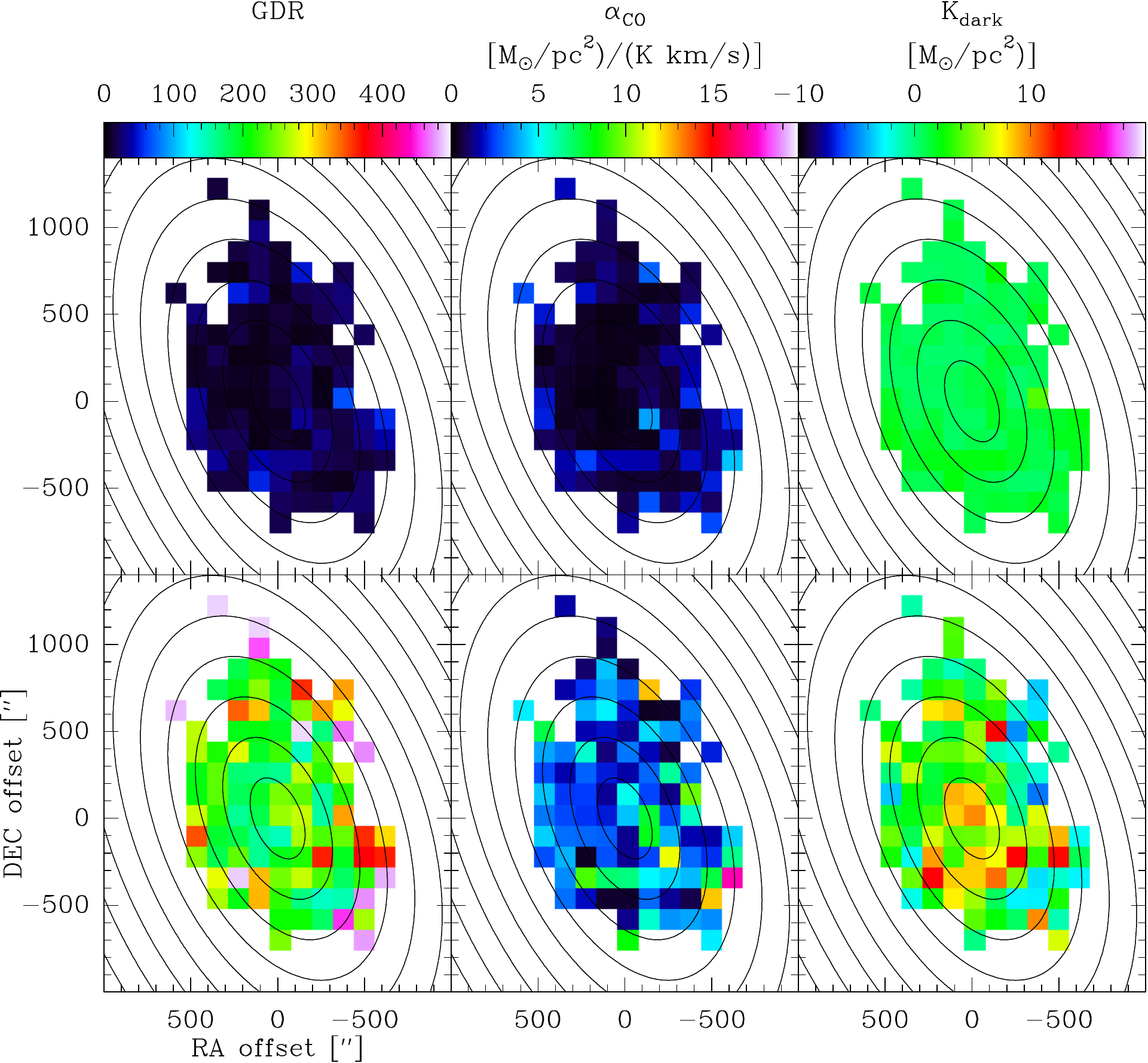} 
	\caption{As for Fig.
	\ref{fig.map_betavar_3sigcut_noGDRprior} but with a 500 cap on \gdr.
	The differences can be seen in the outer parts where some of the high
	\gdr\ pixels from Fig.~\ref{fig.map_betavar_3sigcut_noGDRprior} which
	were white because they had values over 500.}
	\label{fig.map_betavar_3sigcut_GDRprior} 
\end{figure}
}

\newcommand{\FigMapBetavarSigcutzeroGRDlimit}{ 
\begin{figure}
	\centering
	\includegraphics[width=\hsize{}]
	{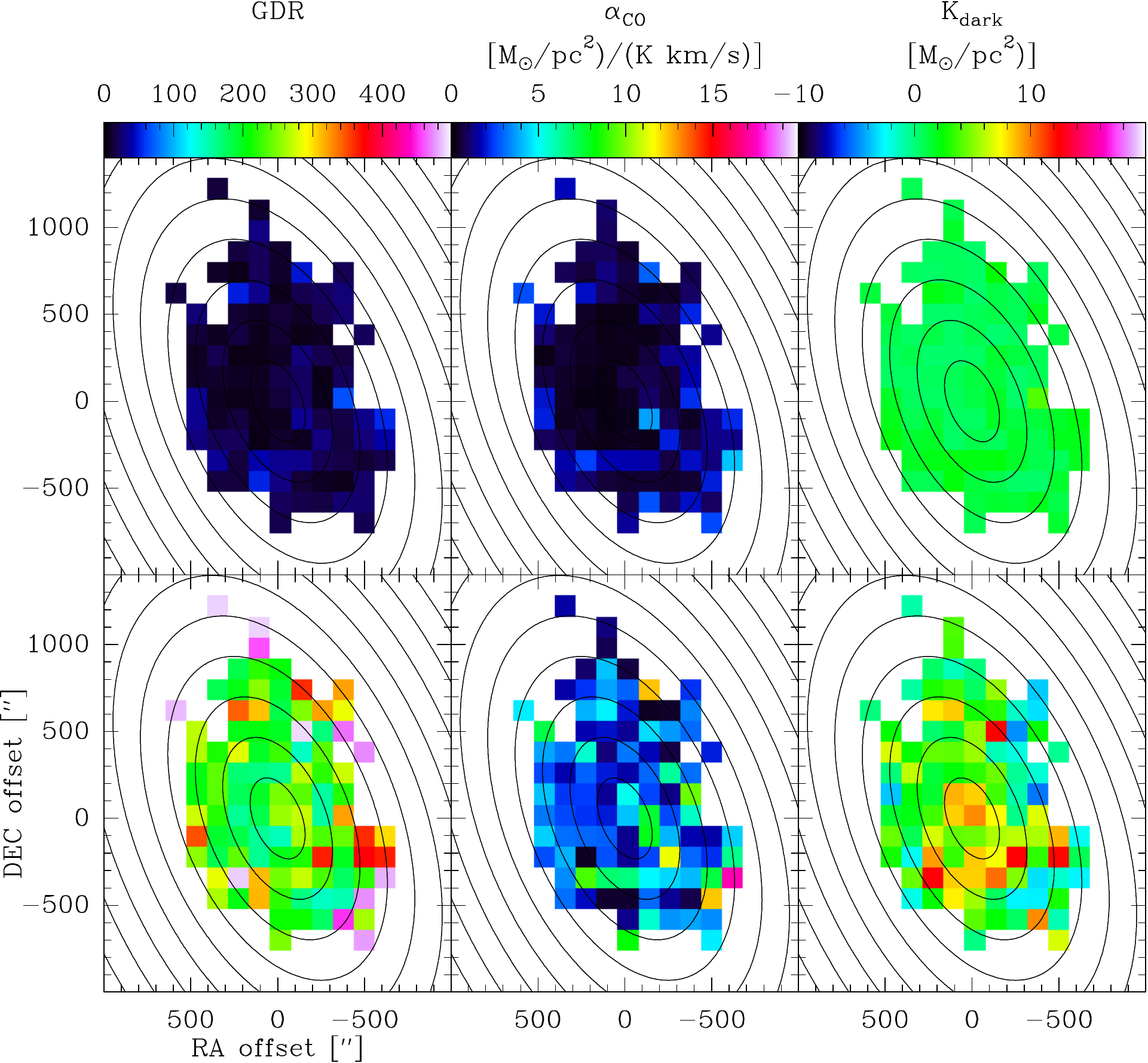} 
	\caption{Results of
	Bayesian analysis with a 0$\sigma$ cut in CO and \gdr\ limited to 500.
	Top row is \kdark\ (left) and uncertainty in \kdark\ (right), both with
	the color scale to the right and in units of $\Msunpsqpc$.
	The second row is $\xco$ (left) and uncertainty in $\xco$ (right), both
	with the color scale to the right and in units of $\Msunpsqpc$ per
	$\Kkms$.
	Bottom row is \gdr\ (left) and uncertainty in \gdr\ (right), both with
	the color scale to the right.
	As with the other figures, we have adopted the variable-beta dust
	surface density shown in Figure 2.}
	\label{fig.map_betavar_0sigcut_GDRprior} 
\end{figure}
}

\newcommand{\FigMapBetavarSigcutthreeNoGRDlimit}{ 
\begin{figure}
	\centering 
	\includegraphics[width=\hsize{}]
	{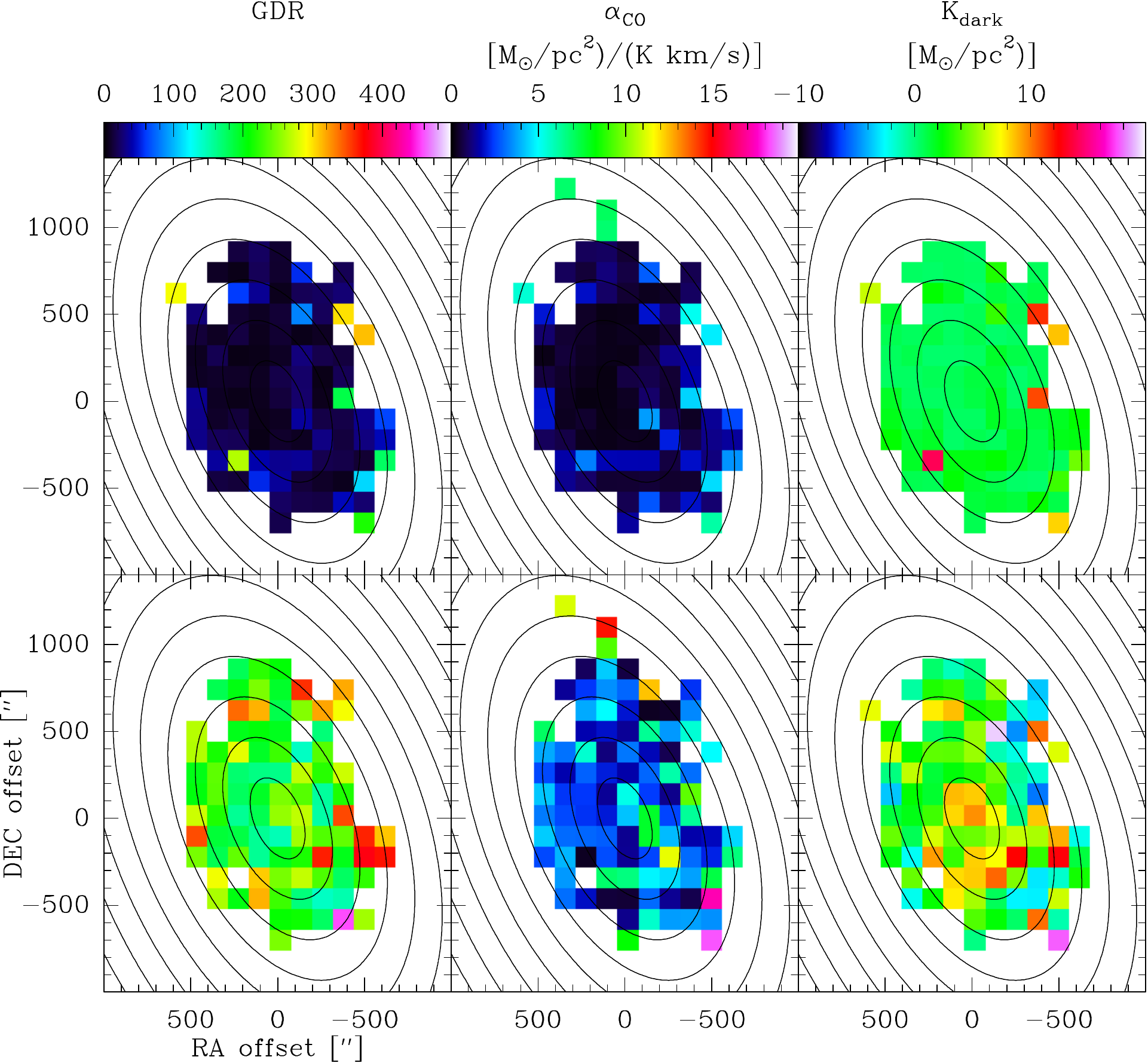} 
	\caption{Results
	of Bayesian analysis with a 3$\sigma$ cut in CO and no cap on \gdr.
	Top row is \kdark\ (left) and uncertainty in \kdark\ (right), both with
	the color scale to the right and in units of $\Msunpsqpc$.
	The second row is $\xco$ (left) and uncertainty in $\xco$ (right), both
	with the color scale to the right and in units of $\Msunpsqpc$ per
	$\Kkms$.
	Bottom row is \gdr\ (left) and uncertainty in \gdr\ (right), both with
	the color scale to the right.
	As with the other figures, we have adopted the variable-beta dust
	surface density shown in Figure 2.}
	\label{fig.map_betavar_3sigcut_noGDRprior} 
\end{figure}
}

\newcommand{\FigRadialBetavarSigcutthreeGRDlimit}{ 
\begin{figure}
	\centering 
	\includegraphics[width=\hsize{}]
	{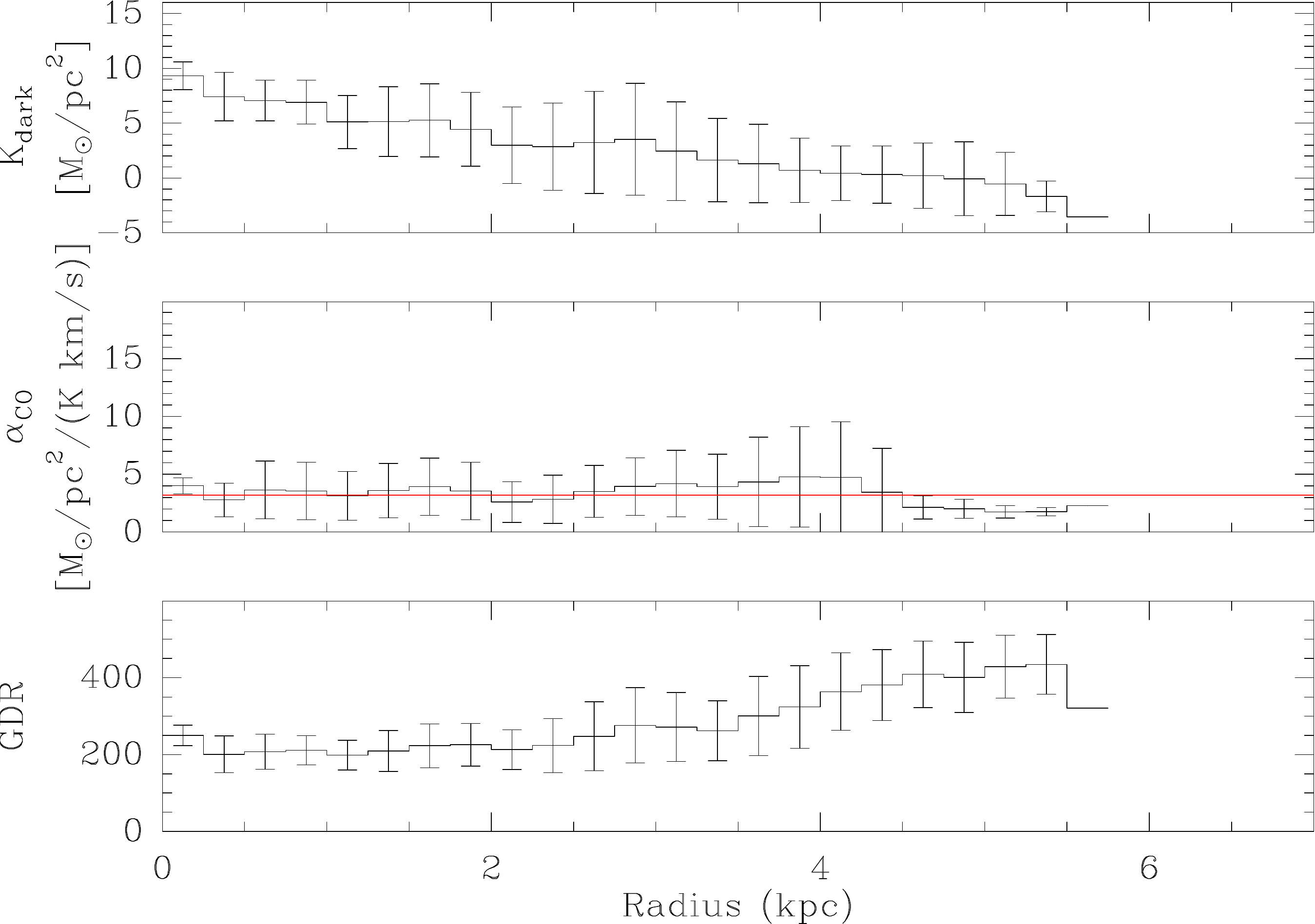} 
	\caption{Radial
	variation of \kdark\ , \aco, and \gdr\ for the simulation with a cut at
	3$\sigma$ for the CO and \gdr\ capped at 500. The value is computed
	using the maps in Fig.~\ref{fig.map_betavar_3sigcut_GDRprior} and
	weighting each macropixel according to is area within a given radial
	annulus. The error bars indicate the dispersion within this ring. }
	\label{fig.radial_betavar_3sigcut_GDRprior} 
\end{figure}
}

\newcommand{\FigRadialBetavarSigcutthreenoGRDlimit}{ 
\begin{figure}
	\centering 
	\includegraphics[width=\hsize{}]
	{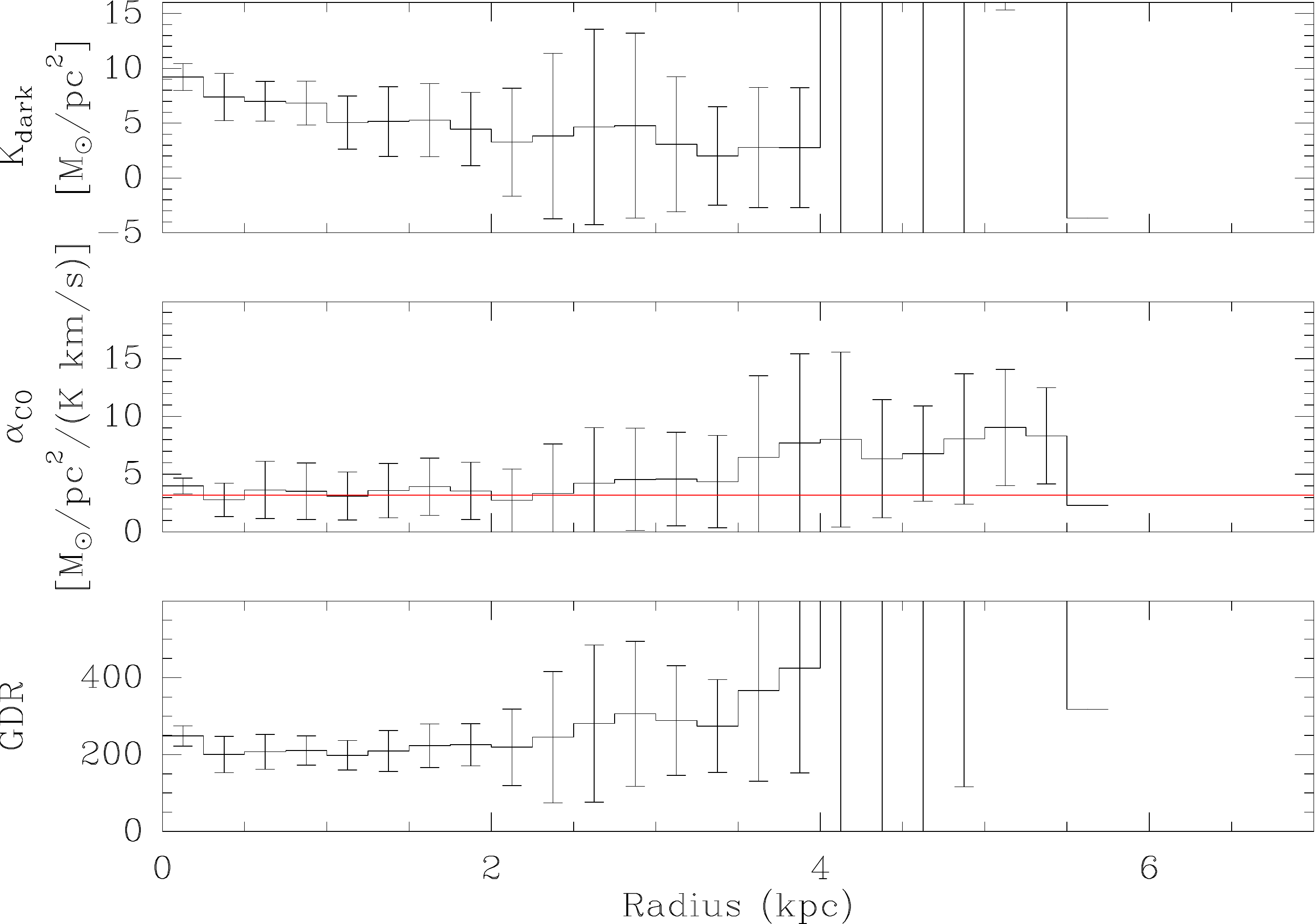} 
	\caption{Radial
	variation of \kdark\ , \aco, and \gdr\ for the simulation with a cut at
	3$\sigma$ for the CO but no cap on \gdr. The value is computed using the
	maps in Fig.~\ref{fig.map_betavar_3sigcut_noGDRprior} and weighting each
	macropixel according to is area within a given radial annulus. The error
	bars indicate the dispersion within this ring. A divergence of \kdark\
	and \gdr\ can be seen in the outer part.}
	\label{fig.radial_betavar_3sigcut_noGDRprior} 
\end{figure}
}

\newcommand{\FigRadialBetavarSigcutzeroGRDlimit}{ 
\begin{figure}
	\centering 
	\includegraphics[width=\hsize{}]
	{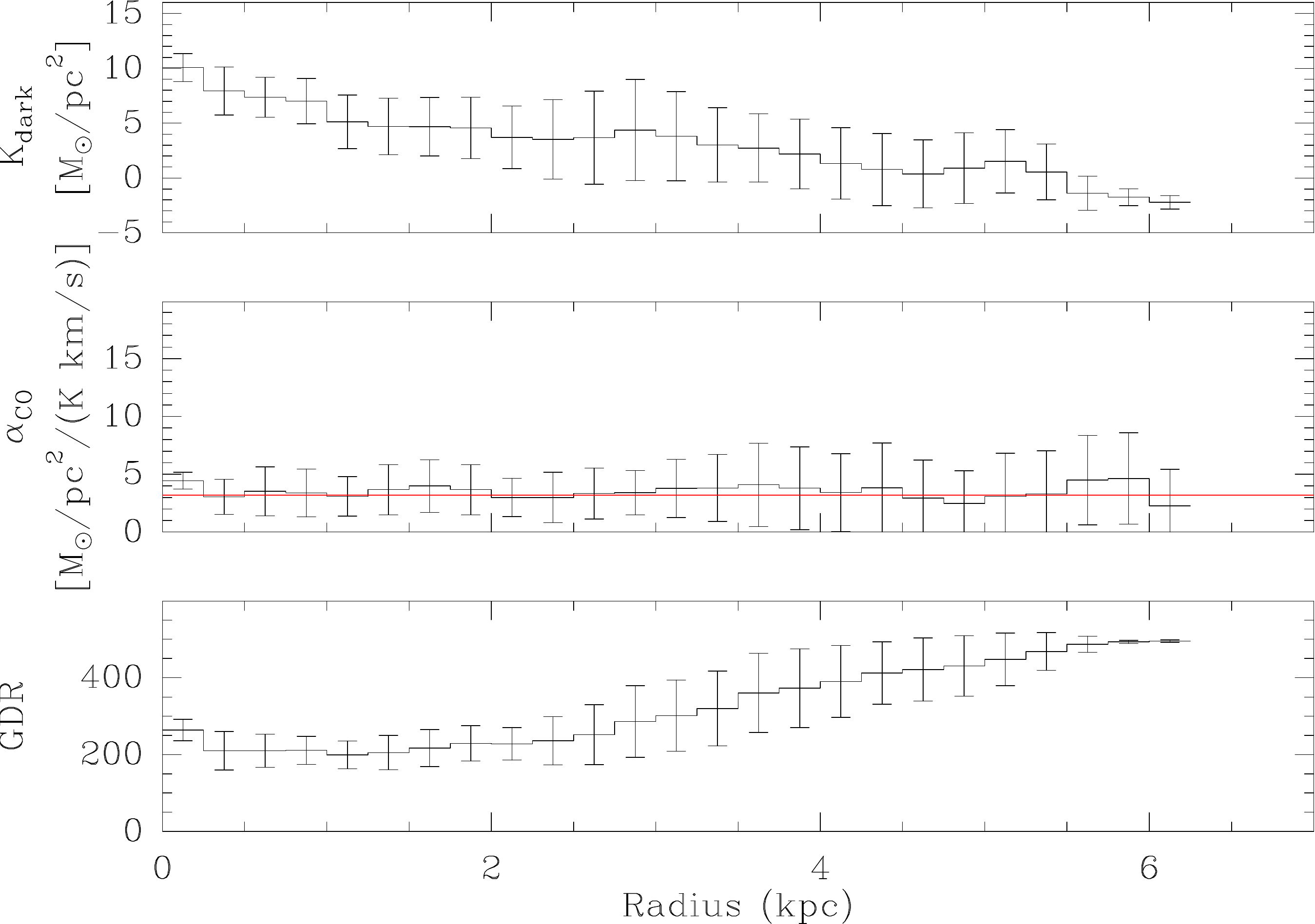} 
	\caption{Radial
	variation of \kdark\ , \aco, and \gdr\ for the simulation with a cut at
	0$\sigma$ for the CO and \gdr\ capped at 500. The value is computed
	using the maps in Fig.~\ref{fig.map_betavar_0sigcut_GDRprior} and
	weighting each macropixel according to is area within a given radial
	annulus. The error bars indicate the dispersion within this ring. The
	comparison with the preceding figures shows that the cap on \gdr\ is
	critical to avoid diverging values of \gdr\ and \kdark.}
	\label{fig.radial_betavar_0sigcut_GDRprior} 
\end{figure}
}

\newcommand{\FigNumpix}{ 
\begin{figure}
	\centering
	\includegraphics[width=\hsize{}]{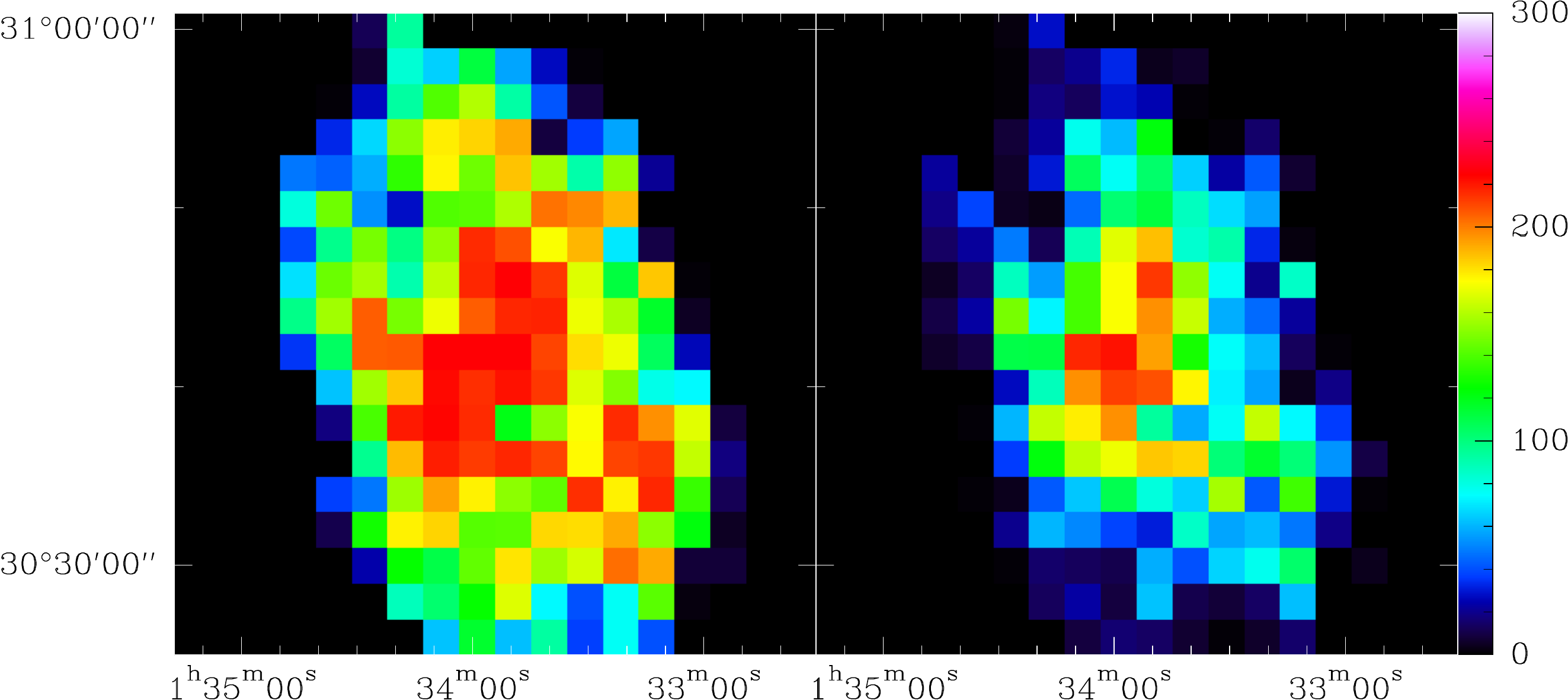} 
	\caption{Maps of the
	number of pixels in each macropixel for the 0$\sigma$ (left) and
	$3\sigma$ (right) CO cuts} 
	\label{fig.numpix} 
\end{figure}
}

\newcommand{\FigCompare}{ 
\begin{figure}
	\begin{centering}
		\includegraphics[trim={0 0 3cm 0cm},
		width=0.45\hsize{}]{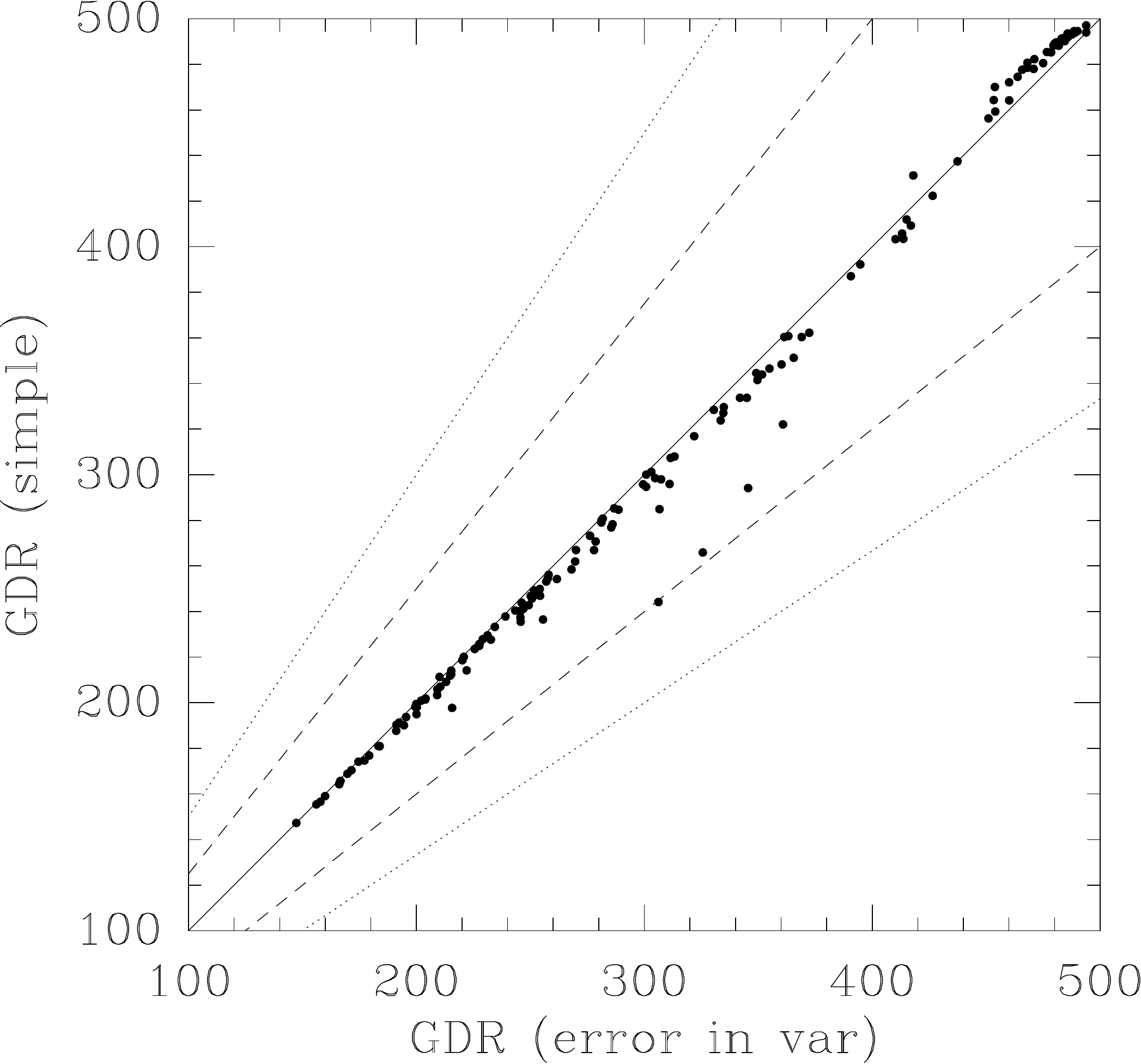} 
		\includegraphics[trim={0 0 3cm
		0cm}, width=0.45\hsize{}]{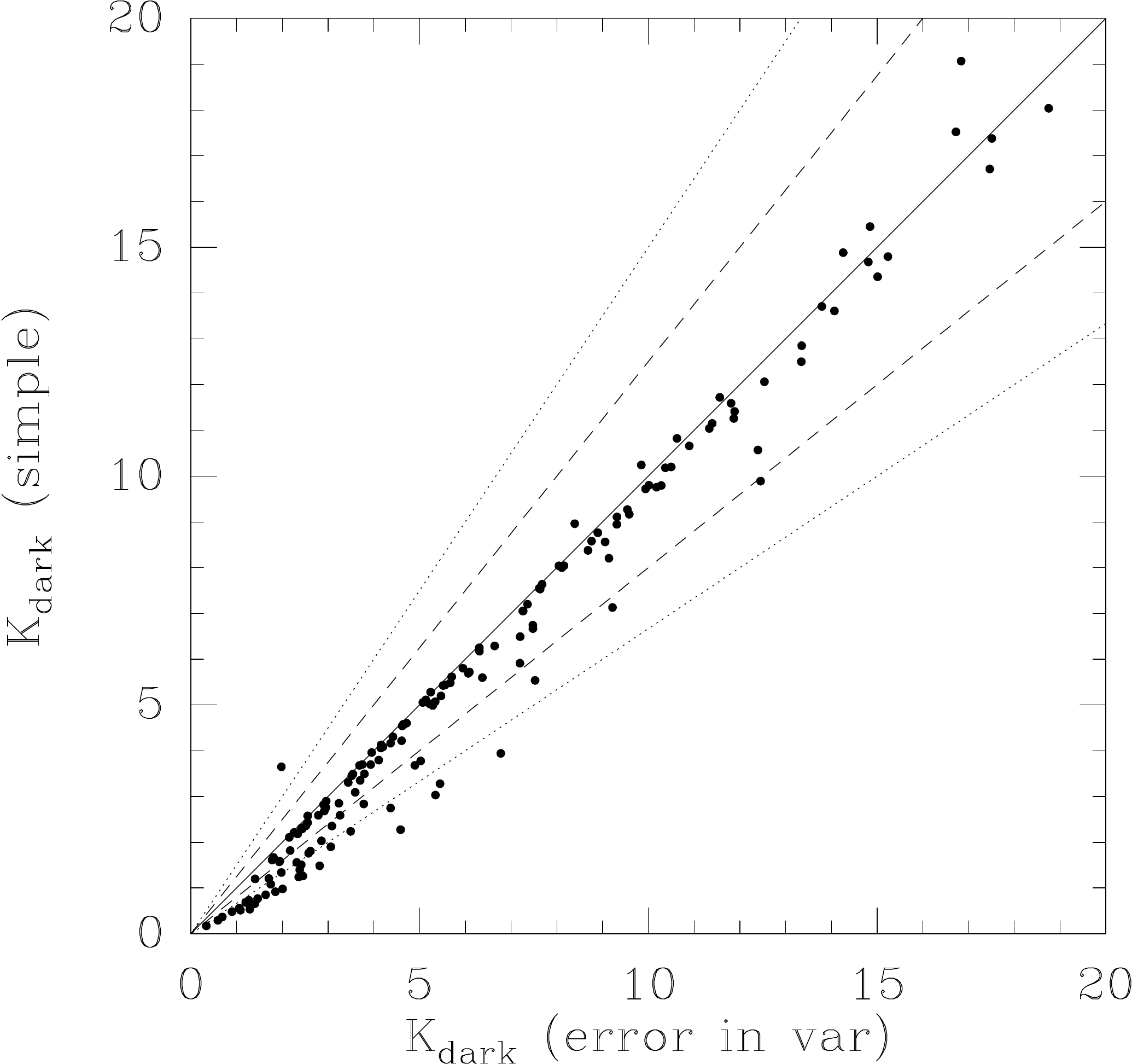} 
		\includegraphics[trim={0 0
		3cm 0cm}, width=0.45\hsize{}]{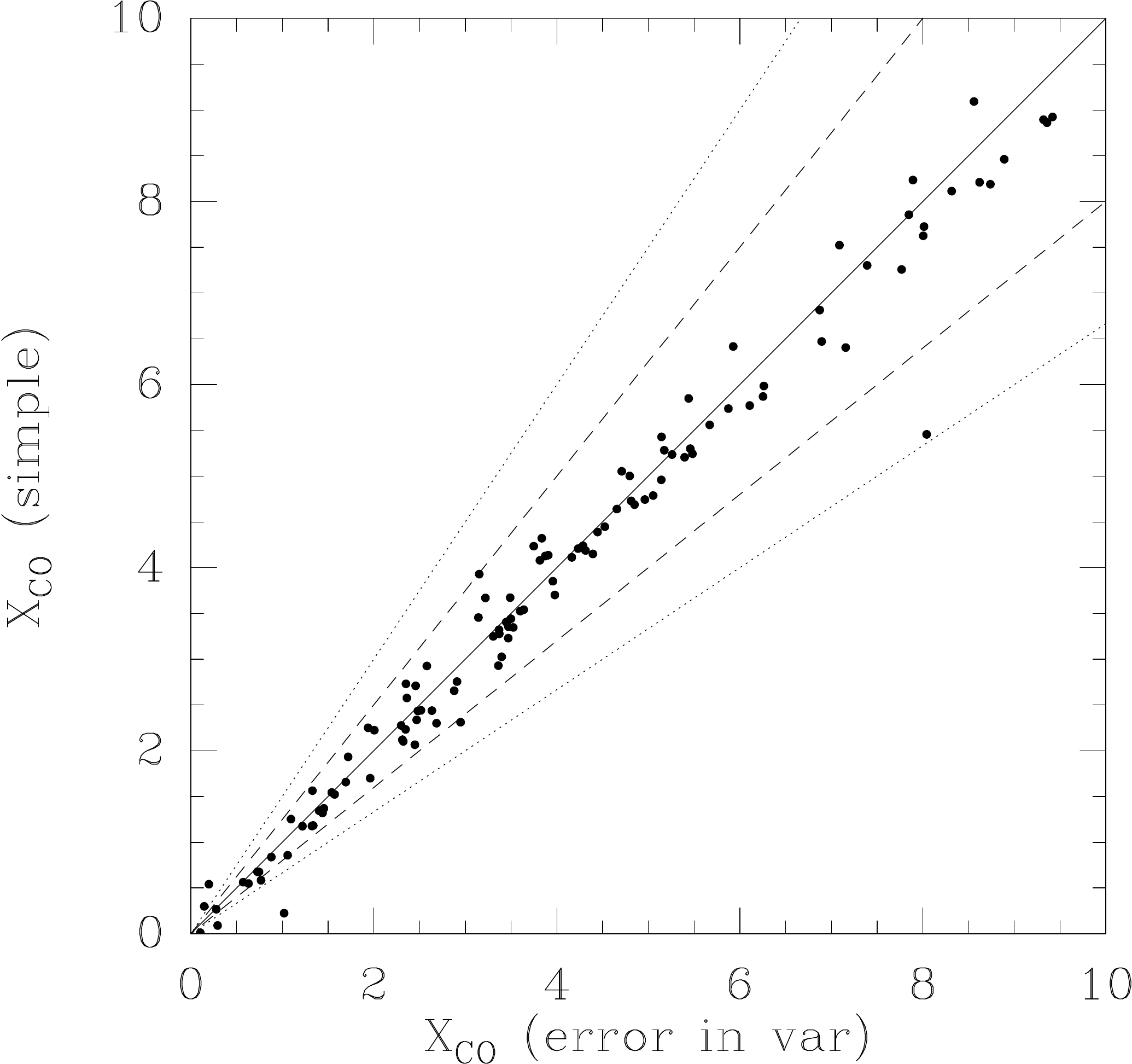} 
	\end{centering}
	\caption{Comparison of rapid and full errors-in-variables Bayesian
	simulations. The solid line represents the equality of the two
	quantities and the dashed (resp. dotted) lines are constant ratios of
	0.25 (resp. 0.5). } 
	\label{fig.compare} 
\end{figure}
}

\newcommand{\FigNHiDust}{ 
\begin{figure}
	\begin{centering}
		\includegraphics[width=0.98\hsize{}]{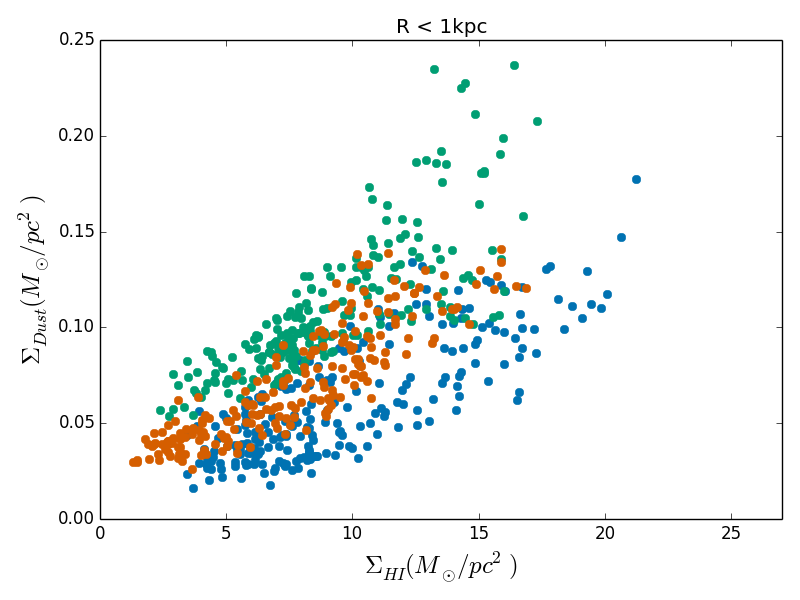}
		\includegraphics[width=0.98\hsize{}]{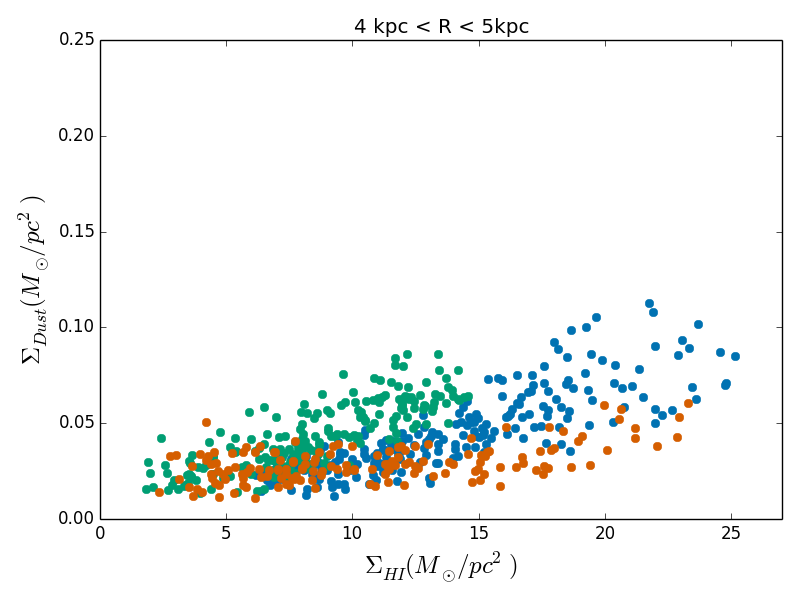}
	\end{centering}
	\caption{{\it top:} Link between \Hi\ column density and
	dust surface density for 3 macro-pixels near the center of M33. Each
	color represents the pixel values of \NHi\ and $\Sdust$ for a single
	macro-pixel. {\it bottom:} Same as above but for 3 macro-pixels between
	4 and 5 kpc from the center.} 
	\label{fig.NHiDust} 
\end{figure}
}

\newcommand{\TabNotation}{ 
\begin{table*}
	\caption{
	\label{tab.notation}Description variables} 
	\begin{centering}
		\begin{tabular}
			{lll} \hline Variable & Quantity & Unit \\
			\hline $\ico$
			& Observed \twCO\Jtwo\ integrated intensity & \Kkms\\
			$\ihi$ & Observed
			\ion{H}{i} 21cm integrated intensity & \Kkms\\
			\NHi & Atomic gas column
			density & \pscm\\
			\NHt & Molecular gas column density & \pscm\\
			$\Shi
			\tablefootmark{a}$ & Atomic gas surface density & \Msunpsqpc\\
			$\Sht
			\tablefootmark{a}$ & Molecular gas surface density & \Msunpsqpc\\
			$\Sgas
			\tablefootmark{a}$ & Total gas surface density & \Msunpsqpc\\
			$\Sdust
			\tablefootmark{a}$ & Dust surface density & \Msunpsqpc\\
			$\sigma_{\Sdust} 
			\tablefootmark{a}$ & Uncertainty on dust surface
			density & \Msunpsqpc\\
			$\aco 
			\tablefootmark{a}$ & Conversion factor from
			$I_{CO}$ to \Ht\ mass surface density & $\unit{\Msun/(pc^2\Kkms)}$\\
			$\xco$ & $\ratioo$ Conversion factor from $I_{CO}$ to \Ht\ column
			density & \Xunit\\
			\gdr\ 
			\tablefootmark{b} & Gas to dust mass ratio &
			unitless \\
			$\kdark 
			\tablefootmark{a}$ & CO dark gas surface density &
			$\unit{\Msun/pc^2}$ \\
			$\kdarkp $ & CO dark gas column density & $\pscm$
			\\
			$\kappa$ & Dust opacity & \unit{cm^2/g}\\
			$\beta$ & Dust emissivity
			index & unitless\\
			$\sigmadust$ & Dust cross section &
			$\unit{cm^{-2}/H}$\\
			$B_{\nu,T}$ & Black body surface brightness at
			frequency $\nu$ and temperature $T$ & \unit{Jy/sr}\\
			\Zgas &
			gas-phase-metal-fraction & unitless\\
			$m_p$ & Proton mass & \unit{g}\\
			\hline 
		\end{tabular}
		\tablefoot{ 
		\tablefoottext{a}{Quantities involving
		masses are considered without taking the helium fraction into account}
		\tablefoottext{b}{\gdr\ is the same quantity as $\delta_{\emr{GDR}}$ in
		\citet{Leroy.2011}} } 
	\end{centering}
\end{table*}
}

\begin{document} 

\title{The molecular gas mass of M33}

\subtitle{}

\titlerunning{The Molecular Gas Mass of M33}

\author{Gratier, P.\inst{\ref{LAB}} 
\and Braine, J.\inst{\ref{LAB}} 
\and
Schuster, K.\inst{\ref{IRAM_fr}} 
\and Rosolowsky, E.\inst{\ref{UA}} 
\and
Boquien, M.\inst{\ref{Antofagasta}} 
\and Calzetti, D.\inst{\ref{UMass}}
\and Combes, F.\inst{\ref{LERMA}} 
\and Kramer, C.\inst{\ref{IRAM_es}}
\and Henkel, C.\inst{\ref{MPIfR},\ref{KaUniv}} 
\and Herpin,
F.\inst{\ref{LAB}} 
\and Israel, F.\inst{\ref{Leiden}} 
\and Koribalski,
B. S.\inst{\ref{CSIRO}} 
\and Mookerjea, B. \inst{\ref{TataIFR}} 
\and
Tabatabaei, F. S.
\inst{\ref{IAC}} 
\and Röllig, M. \inst{\ref{KOSMA}} 
\and van der Tak,
F. F. S. \inst{\ref{SRON},\ref{Kapteyn}} 
\and van der Werf. P.
\inst{\ref{Leiden}} 
\and Wiedner, M.\inst{\ref{LERMA}} } \institute{
Laboratoire d'astrophysique de Bordeaux, Univ. Bordeaux, CNRS, B18N,
allée Geoffroy Saint-Hilaire, 33615 Pessac, France 
\label{LAB} 
\and
Institut de Radioastronomie Millimétrique (IRAM), 300 Rue de la
Piscine, F-38406 Saint Martin d’Hères, France 
\label{IRAM_fr} 
\and
Department of Physics, 4-181 CCIS, University of Alberta, Edmonton, AB
T6G 2E1, Canada 
\label{UA} 
\and Unidad de Astronomía, Fac. Cs.
Básicas, Universidad de Antofagasta, Avda. U. de Antofagasta 02800,
Antofagasta, Chile 
\label{Antofagasta} 
\and Department of Astronomy,
University of Massachusetts—Amherst, Amherst, MA 01003, USA
\label{UMass} 
\and Observatoire de Paris, LERMA (CNRS: UMR 8112), 61 Av.
de l'Observatoire, 75014, Paris, France 
\label{LERMA} 
\and Instituto de
Radioastronoma Milimtrica (IRAM), Av. Divina Pastora 7, Nucleo Central,
E-18012 Granada, Spain 
\label{IRAM_es} 
\and Max-Planck-Institut für
Radioastronomie, Auf dem Hügel 69, 53121, Bonn, Germany 
\label{MPIfR}
\and Astron. Dept., King Abdulaziz University, P.O. Box 80203, Jeddah
21589, Saudi Arabia 
\label{KaUniv} 
\and Leiden Observatory, Leiden
University, PO Box 9513, 2300 RA, Leiden, The Netherlands 
\label{Leiden}
\and CSIRO Astronomy and Space Science, Australia Telescope National
Facility, PO Box 76, Epping, NSW 1710, Australia 
\label{CSIRO} 
\and Tata
Institute of Fundamental Research, Homi Bhabha Road, 400005, Mumbai,
India 
\label{TataIFR} 
\and Instituto de Astrofísica de Canarias, Vía
Láctea S/N, E-38205 La Laguna, Spain; Departamento de Astrofísica,
Universidad de La Laguna, E-38206 La Laguna, Spain 
\label{IAC} 
\and
KOSMA, I.
Physikalisches Institut, Universität zu Köln, Zülpicher Strasse 77,
50937, Köln, Germany 
\label{KOSMA} 
\and SRON Netherlands Institute for
Space Research, Landleven 12, 9747 AD, Groningen, The Netherlands
\label{SRON} 
\and Kapteyn Astronomical Institute, University of
Groningen, The Netherlands 
\label{Kapteyn} } 
\date{}

\abstract{Do some environments favor efficient conversion of molecular
gas into stars? To answer this, we need to be able to estimate the \Ht\
mass.
Traditionally, this is done using CO observations and a few assumptions
but the Herschel observations which cover the Far-IR dust spectrum make
it possible to estimate the molecular gas mass independently of CO and
thus to investigate whether and how the CO traces \Ht. Previous attempts
to derive gas masses from dust emission suffered from biases. Generally,
dust surface densities, \Hi\ column densities, and CO intensities are
used to derive a gas-to-dust ratio (\gdr) and the local CO intensity to
\Ht\ column density ratio (\xco), sometimes allowing for an additional
CO-dark gas component (\kdark).
We tested earlier methods, revealing degeneracies among the parameters,
and then used a sophisticated Bayesian formalism to derive the most
likely values for each of the parameters mentioned above as a function
of position in the nearby prototypical low metallicity ($12 + \log(O/H)
\sim 8.4$) spiral galaxy M33.
The data are from the IRAM Large Program mapping in the CO(2--1) line
along with high-resolution \Hi\ and Herschel dust continuum
observations. Solving for \gdr, $\xco$, and $\kdark$ in macropixels $500
\pc$ in size, each containing many individual measurements of the CO,
\Hi, and dust emission, we find that ($i$) allowing for CO dark gas
($\kdark$) significantly improves fits; ($ii$) $\kdark$ decreases with
galactocentric distance; ($iii$) \gdr\ is slightly higher than initially
expected and increases with galactocentric distance; ($iv$) the total
amount of dark gas closely follows the radially decreasing CO emission,
as might be expected if the dark gas is \Ht\ where CO is
photodissociated. The total amount of \Ht, including dark gas, yields an
average $\xco$ of twice the galactic value of $\sciexp{2}{20}\Xunit$,
with about 55\% of this traced directly through CO. The rather constant
fraction of dark gas suggests that there is no large population of
diffuse \Ht\ clouds (unrelated to GMCs) without CO emission. Unlike in
large spirals, we detect no systematic radial trend in $\xco$, possibly
linked to the absence of a radial decrease in CO line ratios. }


\maketitle{} 
\section{Introduction}

\TabNotation{}

Recent work has shown that large-scale star formation in galaxies is
strongly linked to the molecular gas reservoir, in particular the dense
molecular gas, and less so to the total amount of neutral gas (\Ht\ +
\Hi) \citep{Kennicutt.2012,Lada.2012}. If we are to understand what
affects the relationship between molecular gas and star formation, we
need to be able to measure the amount of molecular gas at all positions
within the disk of galaxies, ideally down to the scale of individual
star-forming regions.
In low-metallicity objects, we are very far from such an understanding.
The cosmic star-formation rate density rises rapidly with redshift
\citep{Madau.2014}, suggesting that either or both the molecular gas
content and the star-formation efficiency (mass of stars formed per unit
time and unit \Ht\ mass) also increase while the fraction of metals
decreases with redshift \citep{Combes.2013}. This is such that what we
learn about local star formation at subsolar metallicities may be useful
to better interpret observations of the young universe.
The small Local Group spiral galaxy M33 has a half-solar metallicity and
is near enough \citep[840\unit{kpc},][]{Galleti.2004} to resolve Giant
Molecular Clouds (GMCs) and has an inclination ($i=56^{\circ}$) that
makes the position of the clouds in the disk well defined (in contrast
to e.g. M31).

The whole bright stellar disk of M33 (up to a radius of $\sim 7 \kpc$)
was recently observed in the CO\Jtwo\ line down to a very low noise
level \citep{Druard.2014,Gratier.2010} using the IRAM 30 meter telescope
on Pico Veleta. The single-dish CO\Jtwo\ data do not suffer from missing
flux problems which is an essential asset to the understanding of the
entire molecular phase in the galactic disk. M33 is a chemically young
galaxy with a high gas mass fraction and as such represents a different
environment in which to study cloud and star formation with respect to
the Milky Way. As the average metallicity is subsolar by only a factor
of two and the morphology remains that of a rotating disk, M33
represents a stepping stone towards lower metallicity and less regular
objects.
Measuring the link between CO and \Ht\ is particularly important given
the evidence that the conversion of \Ht\ into stars becomes more
efficient at lower metallicities
\citep{Gardan.2007,Gratier.2010,Druard.2014,Hunt.2015}.

With the advent of high resolution dust maps in the Herschel SPIRE and
PACS, and Spitzer MIPS and IRAC bands it is possible to determine
reliable dust column densities with spatial resolution close to the size
of individual GMCs in M33 \citep[see][]{Kramer.2010,
Braine.2010a,Xilouris.2012}. Under the assumption of local independence
of the gas-to-dust ratio (\gdr) with respect to the \Ht/\Hi\ fraction,
it is possible to determine the local CO intensity to \Ht\ column
density ratio (\xco).

A simplified global version of such an approach has been applied in
\citet[Fig. 4 of ][]{Braine.2010a}. A more sophisticated method based on
maximizing correlation between dust column density structure and that of
the gas as derived from \Hi\ and CO through an optimal $\xco$ factor has
recently been proposed and successfully demonstrated by
\citet{Leroy.2011} and \citet{Sandstrom.2013}. 

However, these methods have biases and/or degeneracies which will be
studied in Sects.~\ref{sec.NH2ICO} and \ref{sec.LSpresent}, in
particular they often do not consider a possible contribution from CO
dark molecular gas. In this work, the dust, CO, and \Hi\ data covering
the disk of M33 are analyzed using existing these methods along with
simulations to quantify bias and degeneracy. A new Bayesian approach is
then used and tested in order to calculate the \gdr\ and $\xco$ for any
position but also the amount of potential CO dark gas, unseen in \Hi\ or
CO. All the methods take as a basic assumption that any gas not traced
by CO, or potentially optically thick \Hi, contains dust with similar
properties as in the gas traced by CO and \Hi. This is common to all
other studies using dust emission.

\section{Data} \FigDust{}

The CO data are from the recently completed CO\Jtwo\ survey of M33,
which now covers the bright optical disk at high sensitivity
\citep{Druard.2014,Gratier.2010a,Gardan.2007}. The \Hi\ data are from
\citet{Gratier.2010a}. In both lines, we use the datasets produced at
$25''$ resolution. The dust surface density is estimated from the
Herschel observations \citep{Kramer.2010,Boquien.2011,Xilouris.2012},
using the 100, 160, 250, and 350 micron flux densities convolved when
necessary to a resolution of 25$''$ (see Fig.~\ref{fig.dust}). Thus, the
linear spatial resolution at which this study is carried out is 100pc.

In Figure~\ref{fig.dust} (left panel), we show the dust surface density
estimated from the SPIRE 250 and 350\mum\ fluxes, using the ratio of
these two bands to define the temperature, and assuming a dust opacity
of $\kappa = 0.4 (\nu / 250 \GHz)^2$ cm$^2$ per gram of dust
\citep{Kruegel.1994}, or $\kappa_{350} = 4.7 \unit{cm^2 g^{-1}}$ at
350\mum. 

It is now clear that the dust emissivity index, traditionally designated
$\beta$, is not necessarily 2 as has generally been assumed.
In particular \citet{Tabatabaei.2014} have shown that $\beta$ is
variable and lower in M33 ($\beta= 2-1.3$ from the center to the outer
disk). However, without being able to calibrate the value $\kappa$ at
the wavelength of interest, it is difficult to be sure of the constant
(0.4 above for the dust opacity) as extrapolations have generally
assumed $\beta = 2$. If the intrinsic $\beta$ of the dust grains is less
than 2, then using $\beta = 2$ will result in an underestimate of the
temperature and thus an overestimate of the dust mass (compare the two
panels of Fig.~\ref{fig.dust}). In this context, a more accurate but
more complex means of deriving the dust surface density has been tested.
\citet{Tabatabaei.2014} find a link between the galactocentric distance
and $\beta$ in M33 (their Fig.~3). This $\beta(r)$is used to derive dust
temperatures over the disk of M33.

In a similar way as in \citet{Braine.2010a}, we then take pixels with
\Hi\ column density measurements and dust temperatures but no CO
emission and compute the median dust cross-section (\sigmadust) per
H-atom: $\sigmadust = S_{\nu} / (B_{\nu,T} N_H)$, where $S_{\nu}$ is the
dust emission and $B_{\nu,T}$ the Planck black body emissivity for a
frequency $\nu$ and a temperature $T$.
At submillimeter wavelengths the dust emission is optically thin. This
yields a cross-section per H-atom which naturally varies with radius,
much like the metallicity \citep{Magrini.2009}. Using $\sigmadust(r)$,
we calculate the total H (i.e., cold, neutral hydrogen gas:
\Hi\ + \Ht) column density. The dust opacity is $N_H \sigmadust = \kappa
\Sdust= S_{\nu} / B_{\nu,T(\beta)}$, and the dust surface density
$\Sdust = S_{\nu} / (B_{\nu,T(\beta)} \kappa)$. For $\kappa_{350}$ as
above, the dust surface density can be computed for all points in M33,
as shown in Fig.~\ref{fig.dust} (right panel), such that the difference
with respect to Fig.~\ref{fig.dust} (left panel) is that the temperature
is computed with a radially varying $\beta$. The values of $\beta$ are
below 2 in M33 \citep{Tabatabaei.2014} so the temperatures are higher.
Since the Planck function $B_{\nu,T(\beta)} $ increases with T, the dust
surface density in Figure~\ref{fig.dust} (right panel) is lower,
particularly in the outer disk where $\beta$ is lower.

In this work, we only discuss hydrogen content and do not include
helium. As helium is present in both the atomic and molecular phases in
equal proportion, this does not affect the calculations. As in many
other works, we use the term \gdr\ to refer to the hydrogen to dust mass
ratio.

%
%
\section{Dust-derived \Ht\ versus CO intensity} 
\label{sec.NH2ICO}

%
A simple approach is to take the pre-existing map of the \Ht\ column
density based on Herschel and \Hi\ data from \citet{Braine.2010a} where
\NHt\ is estimated from the dust and \Hi\ emission as $\NHt =
(\NHtot-\NHi)/2 $, as in their Figure 4.

In this case, the variables are $\xco$ and, potentially, a CO-dark gas
column density designated \kdarkp. Figure~\ref{fig.nht_ico_nocut_sub}
shows the scatter plots for a sample of three radial bins -- $0\kpc < r
< 1\kpc$, $1\kpc < r < 2\kpc$, and $4\kpc < r < 5\kpc$. These radii show
progressively the transition from an \Ht\ dominated ISM, to approximate
\Hi--\Ht\ equality between radii 1 and 2kpc,\ to the \Hi\ dominated
outer regions.

Thick red lines show the binning of the scatter-plot in 0.5\K\ wide
intervals. The cloud of points are fit by two lines, one assuming $\NHt
= \xco \times \ico$ (light red line) and $\NHt = \xco \times \ico +
\kdarkp$ in green. As described by \citet{Dickman.1986} a $\xco$ ratio
is an average over many different clouds so it cannot be expected to
characterize all clouds, or all of our data points.

\FigNHtICOnocutsub{} 

Figure~\ref{fig.nht_ico_nocut_sub} shows the relationship between the
dust-derived \Ht\ column density and \ico\ for three radial intervals
chosen to represent the inner and outer regions, respectively \Ht\
dominated, slightly \Hi\ dominated ($1-2$ kpc), and strongly \Hi\
dominated with weak CO emission. From the inner to outer regions, the
$\xco$ factor increases, as could be expected given that there is a
metallicity gradient and a decline in CO emission \citep{Gratier.2010a}
and cloud temperature \citep{Gratier.2012b}.

The lines without a \kdarkp\ systematically overestimate the \Ht\ mass
at moderate and high \ico\ and both fits overestimate \NHt\ at high
\ico.
There is no physical reason to expect a constant offset (\kdarkp) but it
appears that there is gas whose dust emission is detected but is not
seen in CO -- this could be optically thick \Hi, molecular gas where CO
has not formed or is photodissociated, low density \Ht\ clouds, or
unexpectedly large quantities of ionized gas.

\section{Leroy-Sandstrom method} 
\label{sec.LSpresent} 
\subsection{Prior
discussion on the gas-to-dust ratio (\gdr)} 
\label{sec.GDRprior} The
\gdr\ is likely well-constrained by the metallicity, at least for
metallicities reasonably close to solar. The solar metallicity is about
$Z = 0.0142$ by mass \citep[][Section 3.1.2]{Asplund.2009}. Assuming the
standard hydrogen-to-dust mass ratio of 100 \citep[][Table
3]{Draine.2007}, the total gas/dust mass ratio is
$M(\Hy+\He+\mbox{gas-phase metals})/M(\mathrm{dust})$, assuming H and He
to be negligible contributors to the dust mass. From Asplund, $M(\Hy) =
0.7154$ and $M(\He) = 0.2703$, and denoting the gas-phase-metal-fraction
as $\Zgas$, we define the hydrogen gas-to-dust mass ratio as $\gdr\ =
(0.7154 + 0.0142\,\Zgas)/(0.0142 (1-\Zgas))$. helium adds just under
40\% to this number. For $\gdr=100$, the typical Galactic value, the
gas-phase-metal-fraction $\Zgas = 0.49$ and 51\% of the metals are in
the dust phase. This value is reasonably robust; for a solar
composition, if $\gdr\ = 100 \pm 20$ then $50 \pm 10$\% of the metals
are in the gas phase.

What about lower metallicity environments? Since dust condenses from the
gas in AGB stellar winds \citep{Gielen.2010} and super nova remnants
\citep{Matsuura.2011}, one expects that when there is less dust and less
metals, the gas-phase metal fraction will tend to be higher. At very low
metallicities, except for very dense environments, the \gdr\ should be
higher than the relation given above due to the difficulty in forming
dust grains and mantles sufficiently quickly such that evaporation or
destruction processes do not reduce the dust mass
\citep{Remy-Ruyer.2014}.

\subsection{Method and application to M33} 
\label{sec.LS}

Developed in \citet{Leroy.2011} and later extended and applied to the
HERACLES/KINGFISH data in \citet{Sandstrom.2013}, the idea is that the
dust emission can be expressed as the sum of the emission from the
atomic and molecular components, implicitly assuming that the
contribution from the ionized gas is negligible. The latter assumption
is likely appropriate and is also common to other studies.
\begin{align}
	\Sgas &= \gdr\ \times \Sdust \notag \\
	&= m_p \times
	\left[ \NHi + 2 \xco \times \ico \right] \notag \\
	&= \Shi + \aco \times
	\ico 
\end{align}

where \aco\ is a surface density conversion factor from \ico\ to \Sht.
Equating the right-hand terms gives us the relation equivalent to
\citet[Eq 3 in ][]{Sandstrom.2013}. In order to allow for some form of
CO dark gas, we allow for an additional term, such that the basic
equation becomes 
\begin{align}
	\Sgas &= \gdr\ \times \Sdust \notag \\
	&=
	m_p \times \left[\NHi + 2 \xco \times \ico + \kdarkp \right] \notag \\
	&= \Shi + \aco \times \ico + \kdark 
	\label{eq.Sgas} 
\end{align}

The procedure is fairly simple: the \aco\ -- \kdark\ space is explored
on a regularly spaced grid and, for each couple (\aco, \kdark), the
dispersion in $\log(\gdr)$ over the ensemble of pixels is computed. The
best fit parameters (\aco, \kdark) are chosen as the ones that minimize
the $\log(\gdr)$ dispersion, similar to what was done in
\citet{Leroy.2011}. \citet[][their appendix]{Sandstrom.2013} later
studied the influence of different methods to identify the best solution
finding robust results over the different methods and settling to using
a minimization of the (robust) standard deviation of the logarithm of
the \gdr. Our maps of M33 cover an area of several thousand beams. This
enables us to look for variations, in particular radial variations, of
\gdr, \aco, and \kdark.
Figure~\ref{fig.scatter_mean_sub} shows this space for three radial
intervals in M33, with a minimum computed assuming that a single value
for each of the three parameters \gdr, \aco, and \kdark\ is appropriate.
The best fits are shown as a function of radius in Figure
~\ref{fig.radial} where the same procedure is applied to concentric
elliptical rings sampling 1 kpc in radius.

\FigtwoDmeansub{} 

\FigRadial{}

%
Figure~\ref{fig.scatter_mean_sub} shows that a very broad region of
\aco\ -- \kdark\ space yields similar quality fits but that a prior on
\gdr\ would help break this degeneracy. The radial behavior shown in
Fig.~\ref{fig.radial} appears somewhat unphysical as the metallicity
gradient necessarily yields an increasing \gdr\ and would be expected to
also yield $\xco$ increasing with radius.

If we assume that $\kdark=0$, then we see from
Fig.~\ref{fig.scatter_mean_sub} (horizontal line where $\kdark=0$) that
the fit is clearly poorer than the best fit. The same is true for the
individual radial bins.
The physical interpretation of \kdark\ is far from straightforward. The
same procedure has been applied but with a filter only accepting pixels
with \ico > $2 \sigma$. The result is essentially the same: the slope of
the ellipses decreases steadily with radius, showing how difficult it is
to measure $\xco$ in the outer regions. The radial variation of the
parameters with radius is shown in Fig.~\ref{fig.radial}. 

The somewhat more complicated nature of the L--S method (3 parameters:
\aco, \gdr\ and \kdark) and the broad degeneracies prompted us to
explore the effect of noise on typical values (Sect. \ref{sec.noise})
and the recoverability of input parameters using realistic simulated
data (Sect. \ref{sec.recov}).

\subsection{Recoverability} 
\label{sec.recov}

In order to check the recoverability of the parameters, we have created
simulated dust observations with known parameters \aco, \gdr\ and
\kdark. The \ico\ and \ihi\ used are the observed values for M33 to
maintain the right correlation between these two quantities. Simulated
observations are created following Eq.~\ref{eq.detdust}. Noise is then
added to each observable quantity \ico, \ihi\ and \Sdust.

%
%
We then create the same figures as in Sect.~\ref{sec.LS}. The figures
are not shown because they are indistinguishable in shape from those in
Section \ref{sec.LS} (Figs.~\ref{fig.scatter_mean_sub} and
Fig.~\ref{fig.radial}).
This is not surprising as the data are the same. However, we can add
many mock runs of the noise and examine how the biases are affected by
differing noise levels and intensity cuts.

Figure \ref{fig.LSCenter} shows the result of 200 sets of trial data
based on the inner kpc. Input parameters are $\xco = 4 \times
10^{20}\pscmpKkms$, $\gdr=150$ and $\kdark=5 \Msunpsqpc$, indicated as
red lines. 

%
It is immediately clear that the optimization (i.e., the lowest
$\log(\gdr)$ dispersion in Fig.~\ref{fig.scatter_mean_sub}) favors
low-valued solutions, with ``optimal'' values clearly below the input.
Even in this high S/N region, $\xco$ is underestimated by 25\% as is
\kdark\ and the \gdr\ by half as much. The \gdr\ is less affected
because the \Hi\ column density is not modified by \kdark\ or $\xco$ but
contributes close to half of the \gdr.

Two variants were tested as well. Although a \kdark\ was present in the
input parameters, we test the values obtained if it is assumed that
$\kdark=0$, as in Eq 3 of \citet{Sandstrom.2013}. In this case, the
\gdr\ is underestimated, presumably because more dust is present (as a
\kdark\ was injected) than what is seen in \Hi\ or CO. Near the center,
(Fig.
\ref{fig.LSCenter}) $\xco$ is underestimated (see middle row) but at
larger radii the situation is different (cf. next paragraph). If
metallicity measurements are reliable, then the \gdr\ is quite
constrained (Sect.~\ref{sec.GDRprior}). The top row shows the values for
$\xco$ and \kdark\ if the true \gdr\ is injected.
If a prior on \gdr\ is injected, then we approximately recover $\xco$
and \kdark. The dispersion in the histograms is rather small, showing
that the results do not depend on the number of realizations.

In the \Hi\ dominated outer regions, Fig.~\ref{fig.LSOuterNoCut} shows
the same biases as before except that $\xco$ is overestimated when
\kdark\ is forced to zero. The prior on \gdr\ again helps recover the
input values with reasonable fidelity. There is only weak CO emission at
these radii so the constraint on $\xco$ is weak. We therefore made a
test excluding values where $I_{co}<2\sigma$. The differences with
respect to the input parameters are somewhat less severe (compare
Figs.~\ref{fig.LSOuterNoCut} and \ref{fig.LSOuterCut}). For the inner
kpc, excluding values below $2\sigma$ makes no difference because
virtually all of the values exceed the threshold.

\FigLSCenter{} \FigLSOuterNoCut{} \FigLSOuterCut{}

\subsection{Noise effects} 
\label{sec.noise}

In order to evaluate the behavior of the Leroy-Sandstrom (L-S) method in
the presence of noise, we took typical values of the CO intensity, the
\Hi\ column, and noise for both, in order to test how the method was
affected by noise. We also allow for the presence of CO dark gas, where
dark means gas not observed in CO or \Hi\ but detected via the emission
of the associated dust. Thus, we start with a single value for each of
$\ico$, $\NHi$ (optically thin assumption), and \kdark\ (dark gas,
assumed constant). Assuming a $\xco$ conversion factor, we calculate the
gas column density ($\NH = 2 \times \xco \times \ico + \NHi +\kdark$)
which we divide by an assumed gas-to-dust ratio (GDR) to obtain a dust
surface density \Sdust, similar to what is estimated from analyses of
Herschel photometric data \citep{Kramer.2010, Xilouris.2012,
Tabatabaei.2014}. We then assume a noise level in the same units for
each of these quantities and generate 1000 samples (value $+$ gaussian
noise) of each of \ico, \NHi, \kdark, and \Sdust.
\Sdust\ after addition of noise is then converted back into a gas
surface density using the same \gdr. The final step is to test a grid of
\xco\ and \kdark\ values, minimizing the sum of 
\begin{equation}
	\left(
	\Sdust\gdr - \aco \ico - \Shi -\kdark \right)^2 
\end{equation}
where the
quantities are after addition of noise and the sum is over the 1000
samples.

The fiducial model has $\ico = 1 \pm 0.25$ K\kms, \NHi$ = 8 \pm 1 \times
10^{20} \pscm$, and $\kdark = 1 \pm 0.25 \times 10^{20} \pscm$ and we
assume the uncertainty in the dust surface density is 25\%. We inject
$\xco = 4 \times 10^{20} \pscmpKkms$ in order to calculate \Sgas\ --
this, along with \kdark, is what we try to get out of the simulations.
The \gdr\ is transparent in that it is used to convert \Sgas\ into
\Sdust\ but then back into \Sgas\ after addition of noise so it
disappears.

Figure~\ref{fig.LSnoise} shows the typical degeneracy between the \xco\
and \kdark\ parameters. The color scale shows the quality of the fit
(the lower the better) and contours show the acceptable regions.
The black dotted lines indicate the average gas-to-dust ratio for the
pixel (i.e., averaged over the 1000 samples for the (\ico,\kdark)
combination). The dotted lines indicate, from left to right, \gdr s of
100, 150, 200, and 250. For this example, with $\ico = 0.5 \pm 0.25
\unit{K\kms}$, the apparently optimal fit is quite far from the input
parameters. These values are quite typical of a large number of the
pixels in M33. \FigLSnoise{}

Figure~\ref{fig.LSimu}a-f show how the retrieved values of \xco\ and
\kdark\ vary with the CO intensity (before adding noise) and the noise
level of the CO observations. The first two figures show the results for
$\NHi = 8 \pm 1 \times 10^{20} \pscm$ and a 25\% uncertainty in the dust
surface density. The second set of figures shows how the recovered \xco\
and \kdark\ values depend on the CO intensity and uncertainty in the
case where $\NHi = 4 \pm 1 \times 10^{20} \pscm$. In the third set,
$\NHi = 8 \pm 1 \times 10^{20} \pscm$ but the uncertainty in the dust
(and thus gas) surface density has been reduced to 10\%.

\FigSimu{}

The result is striking: in all cases, the \xco\ conversion factor and
the $\kdark$ surface density are well recovered for the high CO
intensities and small errors but where the intensity or the S/N is lower
the recovered \xco\ decreases systematically and the amount of dark gas
increases rapidly. A general tendency is seen towards high $\kdark$ and
low \xco\ as the S/N ratio decreases, similar to
Figure~\ref{fig.LSnoise}.

\section{Bayesian method} 
\subsection{Principles} 
\label{sec.bayes_mod}

This method enables us to take into account the uncertainties in all of
the observed quantities and recover the best estimates of the \gdr,
$\xco$, and \kdark\ values. This is done in the Bayesian framework of
errors in variables.

The generative model is defined as:
\begin{align}
	\ihii^{obs} &\sim \mathcal{N}(\ihii^{true}, \sigma_{\ihi,
	i}) 
	\label{eq.distHI}\\
	\icoi^{obs} &\sim \mathcal{N}(\icoi^{true},
	\sigma_{\ico,i}) 
	\label{eq.distCO} \\
	\Sdusti^{true} &= \frac{1}{GDR}
	(\ahi\ihii^{true} + \aco\icoi^{true} + \kdark) 
	\label{eq.detdust} \\
	\Sdusti^{obs} &\sim \mathcal{N}(\Sdusti^{true}, \sigma_{\Sdust})
	\label{eq.distdust} 
\end{align}

The above notation means that the quantity $\ihii^{obs}$ observed at
pixel $i$ has a gaussian distribution centered on the true
$\ihii^{true}$ integrated intensity with a dispersion equal to the
observational uncertainty $\sigma_{\Hi,i}$. Same for the CO in
Eq.~\ref{eq.distCO}. The third line states that the true dust surface
density \Sdusti\ is a function of the true \ihii\ and \icoi\ and the
three model parameters \aco, \gdr\ and \kdark. We assume that the \Hi\
emission is optically thin such that $\XHi =
\sciexp{1.823}{18}\pscmpKkms$ which converted into units of solar masses
per square pc gives $\ahi = 0.0146\unit{M_{\sun}/pc^2/(\Kkms)}$). The
fourth equation states that the observed dust surface density (left) has
a gaussian distribution centered on the true \Sdusti\ with dispersion of
$\sigma_{\Sdust}$. We note that the only equality is for the true
quantities, not the observations.
This method provides an estimate for the {\it true} values of \Sdust,
\ico, and \ihi, as well as the parameters \aco, \gdr, \kdark.

Because the observations are independent, we can express the likelihood
of the parameters knowing the full dataset as the product of the
likelihoods of the parameters knowing each individual datapoint. For $N$
observations, 
\begin{align}
	L(a,b,c,\{\icoi^{true}\},\{\ihii^{true}\},\sigma_{dust}|D) = \notag \\
	p(D|a,b,c,\{\icoi^{true}\},\{\ihii^{true}\},\sigma_{dust}) = \notag \\
	\frac{(2\pi)^N}{\prod_{i=1}^N \sqrt{\sigma_{\ico,i}^2 \sigma_{\ihi, i}^2
	\sigma_{\Sdust}^2} } \notag\\
	\times \prod_{i=1}^N \exp \left[ -
	\frac{(\icoi^{obs}-\icoi^{true})^2}{2 \sigma_{\ico,i}^2} \right] \notag
	\\
	\times \prod_{i=1}^N \exp \left[ - \frac{
	(\ihii^{obs}-\ihii^{true})^2}{2 \sigma_{\ihi, i}^2} \right] \notag \\
	\times \prod_{i=1}^N \exp \left[ - \frac{ (\Sdust^{obs} - a \ihii^{true}
	- b \icoi^{true} - c )^2}{2 \sigma_{\Sdust}^2} \right]
	\label{eq.likelihood} 
\end{align}

where $D$ is the observed dataset $\{\{\icoi^{obs}\},\{\ihii^{obs}\},
\{\Sdusti^{obs}\}\}$ , $a = \ahi/\gdr$, $b=\aco/\gdr$, $c= \kdark/\gdr$.
The likelihood is thus the probablility of having an observed set of
$\{\{\icoi^{obs}\},\{\ihii^{obs}\}, \{\Sdusti^{obs}\}\}$ (i.e., the
observed map of \Sdust, \ico\ and \ihi) given a set of values for
$\ahi/\gdr$, $\aco/\gdr$, $\kdark/\gdr$, \{$\icoi^{true}$\},
$\{\ihii^{true}\}$, and $\sigma_{\Sdust}$. We know the uncertainty in
the \ico\ and \ihi\ observations ($\sigma_{\ihi}, \sigma_{\ico}$) and
the values are input to the calculation. On the other hand, we do not
have a good estimate of the uncertainty in the dust surface density
$\sigma_{\Sdust}$ so this is left as a free parameter and becomes an
output of the calculation.
This $\sigma_{\Sdust}$ will also parameterize Gaussian scatter around
the true relationship so $\sigma_{\Sdust}$ may be larger than the
measurement error, but accounts for additional scatter in the data
\citep{Hogg.2010}.

Thus, there are $4+2N$ parameters ($\ahi/\gdr$, $\aco/\gdr$,
$\kdark/\gdr$, $\sigma_{\Sdust}$ and the $\icoi^{true}$ and
$\ihii^{true}$ for each of the N pixels) to the model and a total of 3N
observations ($\Sdusti^{obs}, \icoi^{obs}, \ihii^{obs}$ for each pixel).

%
Since we are interested in the distribution of the parameters and the
likelihood is a probability distribution for the observations, we use
the Bayes formula to convert from one to the other.
\begin{align}
	p(a,b,c,&\{\icoi^{true}\},
	\{\ihii^{true}\},\sigma_{\Sdust}|D) \propto \notag \\
	&
	p(a,b,c,\{\icoi^{true}\},\{\ihii^{true}\},\sigma_{\Sdust}) \times \notag
	\\
	& p(D|a,b,c,\{\icoi^{true}\},\{\ihii^{true}\},\sigma_{\Sdust})
	\label{eq.posterior} 
\end{align}

The left hand side of Eq.~\ref{eq.posterior} is the posterior
distribution -- the distribution function of the parameters given the
observations. The first term on the right is the prior distribution of
the parameters. In our case, very little information is injected because
only unreasonable values are not tested.
\gdr\ is varied either uniformly from 0 to 500 or uniformly from 0 to
50000, the first case enables to check the influence of using a
physically justifies prior, namely that the \gdr\ values cannot be
higher than 500. $\aco$ is varied from 0 to 30 times \Msunpsqpc/\Kkms
and \kdark\ from -10 to 30 \Msunpsqpc. The $\ihii^{true}$ and
$\icoi^{true}$ parameters are varied between the minimum and the maximum
of the observations. The last term is the probability defined above. The
posterior distribution is explored using an Monte Carlo Markov Chain
(MCMC) code, specifically the {\tt EMCEE} Python implementation
\citep{Foreman-Mackey.2013} of Affine Invariant Ensemble Sampler
described in \citet{Goodman.2010}.

The priors can be summarized as:
\begin{align}
	\gdr\ &\sim \mathcal{U}(0,500) \mbox{ or }
	\mathcal{U}(0,50000) \notag\\
	\aco &\sim \mathcal{U}(0,30)\notag\\
	\kdark &\sim \mathcal{U}(-10,30)\notag\\
	\icoi^{true} &\sim
	\mathcal{U}(\mathrm{min}(\{\icoi^{obs}\}),\mathrm{max}(\{\icoi^{obs}\}))
	\notag\\
	\ihii^{true} &\sim
	\mathcal{U}(\mathrm{min}(\{\ihii^{obs}\}),\mathrm{max}(\{\ihii^{obs}\}))
\end{align}
where $\mathcal{U}(x_{min},x_{max})$ stands for a uniform
distribution between values $x_{min}$ and $x_{max}$.

\subsection{Validation of the Bayesian method} 

To test the Bayesian method, we simulated a dataset using $\aco=2 \times
3.2 \aunits$ (twice the galactic value), $\gdr=150$, $\kdark=10
\Msunpsqpc$ and $\sigma_{\Sdust}=0.01 \Msunpsqpc$.
Since we need to input ``true'' values of \ico\ and \ihi\ in order to
see if we can recover the parameters we inject the observed values of
\ico\ and \ihi. These values are then used to create the ``true'' dust
map as per Eq~\ref{eq.detdust}. Since the method starts with
observations, we take the simulated observed values to be the real
observed values plus noise.
Thus, the calculation uses somewhat noisier values than the real data.
The tests use datapoints (\ihi,\ico) characteristic of the inner disk of
M33. Noise is also added to the ``true'' dust surface density map
(created via Eq~\ref{eq.detdust}).

This model dataset is then used as input into the Bayesian method
described in Sect.~\ref{sec.bayes_mod}. Figure~\ref{fig.BayeTest} shows
the number density of points in the six planes mixing the four
parameters \aco, \gdr, \kdark, and $\sigma_{dust}$ in grayscale. The
orientation of the contours illustrates any degeneracies in the
relationship between the parameters. The input parameters to the
simulation are shown as solid blue lines. The 4 histograms show the
entire set of values for each parameter and the dashed lines show the
median and the $\pm 1\sigma$ and $\pm 2\sigma$. The results contain no
obvious bias and are very close to the input parameters.
Furthermore, the confidence intervals ($\pm 1\sigma$ and $\pm 2\sigma$)
are determined in a self-consistent way.

Figure~\ref{fig.BayeTest} is the result of a simulation of the inner kpc
of M33. In the outer parts, the CO emission is very weak and the gas
(and thus dust) surface density is dominated by the \Hi. The Bayesian
method as proposed here is not always able to measure the $\xco$ factor
where there is little CO emission but an upper limit comes out
naturally. On the other hand, \gdr\ can be measured because \Hi\ is
present in many pixels.

\FigBayesTest{} 

\subsection{Application to M33} 

M33 was divided into 324 macropixels measuring 500pc $\times$ 500 pc,
each containing 225 independent pixels of \Hi, CO, and dust data. This
size is large enough that the parameters are well-defined but small
enough not to be affected by large-scale gradients. From the results for
the macropixels, it is possible to estimate the radial variation of each
parameter. The large number of pixels and macropixels results in an
extremely high computation time -- about six months CPU using a machine
with 12 processors and 128Gb of memory.

Nearly all (99\%) of this time is taken up by the ``error in variables''
approach (using the full model consisting of all four
Eqs.~\ref{eq.distHI} to \ref{eq.distdust}). Thus, given the prohibitive
CPU time, we tested the Bayesian estimation without the error in
variables (using a restricted model consisting of only
Eqs.~\ref{eq.detdust} and \ref{eq.distdust}), which runs in a day so we
can test different hypotheses.
The cases we would like to test are: using the two different dust maps,
with different cuts in CO intensity, and with or without limits on the
value that \gdr\ can take.

The ``error in variables'' approach produces slightly lower
uncertainties but essentially the same values for the parameters \gdr,
$\xco$, and $\kdark$.
This can be seen in Fig.~\ref{fig.compare} which shows the values of
$\kdark$, \aco, and \gdr\ for the error-in-variables and the rapid
versions. In these simulations, the dust surface density for the
variable-$\beta$ was used, only pixels with CO intensities above
3$\sigma$ were included, and \gdr\ was allowed to take values between 0
and 500 (5 times the Galactic value).
\FigCompare{} 

Therefore, we use the rapid (1 CPU-day) computations in the following.

Even with the Bayesian approach, some degeneracy is present. In
Fig.~\ref{fig.map_betavar_3sigcut_noGDRprior} (result) and
\ref{fig.radial_betavar_3sigcut_noGDRprior} (radial), we show the
results for variable-$\beta$ dust with a 3$\sigma$ CO cut but without
placing a limit on \gdr. Both $\kdark$ (upper panel) and \gdr\ (lower
panel) diverge at large radii, where the CO becomes less of a
constraint. This is due to some pixels reaching arbitrarily high \gdr\
values (thousands). If the CO cut is reduced to $0\sigma$, then $\kdark$
and \gdr\ diverge at lower radii. The hydrogen mass to dust mass ratio
in the Milky Way is about 100, close to 140 if He is included.
We thus decided to limit \gdr, not allowing it to go above 500 (close to
700 if He is included). Presumably this is well above any true \gdr\
value for a half-solar metallicity galaxy. The $\xco$ factor is not very
affected by the divergence of $\kdark$ and \gdr\ although it is
difficult to be confident of its value where there is little CO.

\FigMapBetavarSigcutthreeNoGRDlimit{}

\FigRadialBetavarSigcutthreenoGRDlimit{}

Figure~\ref{fig.numpix} shows the maps of the number of measurements
used for each of the macropixels for the 0$\sigma$ and $3\sigma$ CO
cuts.

\FigNumpix{}

Figure~\ref{fig.radial_bayes_compare} is similar to
Fig.~\ref{fig.radial} in that it shows the influence of the choice of
the dust emissivity index $\beta$ on the derived parameters. For the
Bayesian method, as for the LS method, the results are consistent for
\aco\ and \kdark\ but differ for \gdr\ with smaller values found for the
$\beta=2$ dust maps. This is expected as the $\beta=2$ maps has hight
dust surface densities, particularly at higher radii.

\FigRadialBayesCompare{}

\FigMapBetavarSigcutthreeGRDlimit{}

\FigRadialBetavarSigcutthreeGRDlimit{}

\FigMapBetavarSigcutzeroGRDlimit{} 

\FigRadialBetavarSigcutzeroGRDlimit{}

Figures~\ref{fig.map_betavar_3sigcut_GDRprior} (result) and
\ref{fig.radial_betavar_3sigcut_GDRprior} (radial) show the same as
Fig.~\ref{fig.map_betavar_3sigcut_noGDRprior} and
\ref{fig.radial_betavar_3sigcut_noGDRprior} but when \gdr\ cannot take
values above 500. This essentially avoids finding an optimal result at
extremely high \gdr\ and $\kdark$. Where the CO is present in a
significant number of pixels (Fig.~\ref{fig.numpix}), the limitation (of
\gdr) is unnecessary but when the equation really only equates \gdr\ and
$\kdark$ then they are highly degenerate.

Figures~\ref{fig.map_betavar_0sigcut_GDRprior} (result) and
\ref{fig.radial_betavar_0sigcut_GDRprior} (radial) show the radial
variation of $\kdark$, $\xco$, and \gdr\ for the 0$\sigma$ and $3\sigma$
CO cuts.
The similarity shows that when \gdr\ is not allowed to take unphysical
values, the CO cut is not critical.

The values of \gdr\ we find in the outer regions using the
variable-$\beta$ approach are actually consistent with the \gdr\ found
by \citet{Gordon.2014} in the Large Magellanic Cloud. The LMC is a
useful comparison as it is only slightly smaller, less massive, and less
metallic than M33 but the LMC is much more irregular. 

Several interesting features are present. First of all, even though
\gdr\ increases with radius, $\kdark$ decreases. This shows that the
increase in $\kdark$ seen without the limit on \gdr\ was {\it only} due
to the divergent pixels. The $\xco$ shows no clear radial trend. This is
probably unlike large spirals like our own, where a number of works have
suggested the $\xco$ increases with radius
\citep{Sodroski.1995,Braine.1997}, with a particularly low value in the
central regions. However, large spirals also show systematic decreases
in the CO(2--1)/CO(1--0) ratio whereas M33 does not \citep{Druard.2014}.
The value of $\xco$ is only 10\% greater than the Galactic value,
indicated by a horizontal line in Figs.
\ref{fig.radial_betavar_3sigcut_noGDRprior},
\ref{fig.radial_betavar_3sigcut_GDRprior}, and
\ref{fig.radial_betavar_0sigcut_GDRprior}. This may appear surprising as
the $\xco$ factor is expected to increase as the metallicity decreases.

The $\xco$ factor derived here is {\it not} directly comparable to the
values for the Galactic $\xco$ derived using dust and/or gamma-ray
observations because these calculations did not allow for dark gas and
thus attributed all gas (including any CO dark gas) not identified as
\Hi\ to \Ht\ in order to calculate $\xco$. In order to calculate a
comparable ratio, we can add the CO dark gas to the \Ht\ column computed
as $\ico \times \xco$. While typically modeled as a constant, $\kdark$
is not physically expected to be constant as ($a$) \Hi\ is expected to
be optically thick only over very small areas and ($b$) GMC edges, where
H is molecular but CO photodissociated, are only expected to be
associated with GMCs, which occupy a very small fraction of the disk
\cite{Druard.2014}. Thus, we can either take the value of $\kdark$
derived for the CO detected (0 or $3\sigma$) positions in the macropixel
as representative of all positions, or we can assume that the value of
$\kdark$ derived for the CO detected pixels are only valid for those
pixels and assume zero elsewhere. In this way, we may be able to place
upper and lower limits to the total $\xco$ values in M33, including dark
gas.

We thus consider Fig. 10 from \citet{Druard.2014} and uncorrect for
inclination, uncorrect for He, and rescale to a $\xco$ value of 1.1
Galactic -- this is equivalent to dividing their values by 1.24. To
this, we can add the $\kdark$ as computed either in ($a$) or ($b$)
above.

Expressing the CO-emitting \Ht\ and $\kdark$ as surface densities in
Figure~\ref{fig.XK}, it is interesting to note that they are very
comparable for a $\xco= 1.1 X_{gal}$ where $X_{gal}$ is taken to be $2
\times 10^{20} \Xunit$. If we assume that the dark gas is actually
molecular gas, then the two columns should be added in order to compare
with the Galactic $\xco$ factors based on dust or gamma-rays.
Depending on whether $\kdark$ is assumed to be present everywhere at the
level derived from the positions respecting the CO threshold or only for
those positions, the total $\xco$ (dark \Ht\ + CO-emitting \Ht\ divided
by \ico) is about twice Galactic with very little radial variation.
(except for the case where the only pixels with $\kdark$ are those above
$3\sigma$ in CO). The uncertainties increase dramatically beyond 4.5 kpc
so we have not been able to derive constraints for the very outer disk
of M33.

\FigXK{}

Although we initially expected \kdark\ to increase (at least with
respect to CO) with galactocentric distance as in \citet[Fig. 15 of
][]{Pineda.2013}, is not surprising the \kdark\ decreases with radius
because the UV field decreases much more quickly than the metallicity.
As for the expected increase of \Xco\ with galactocentric radius as is
observed in large spirals \citep{Sodroski.1995, Braine.1997}, it is not
seen in M33. This was initially a surprise but the constant CO line
ratios (2--1/1--0 \emph{and} 3--2/1--0) support this. In large spirals
we see clear decreases in these line ratios (and increases in \Xco), but
this is not the case in M33.
We did not initially expect \kdark\ to follow the CO column density
variation -- that came out of the analysis.
However, it is natural if the CO dark gas is in the outskirts of GMCs.
This implies that there is no large population of diffuse \Ht\ clouds
(unrelated to GMCs) without CO emission.

Our findings are in apparent disagreement with \citet{Pineda.2013}.
However, \citet{Pineda.2013} computed H column densities assuming a
constant $\ratio$ ratio. Introducing a radially decreasing $\ratio$
would at least reduce the difference in our findings. Our findings are
in agreement with \citet{Mookerjea.2016} who find more CO dark gas near
the center than in the BCLMP302 region, although it is very difficult to
generalize from a small number of regions. While we describe \kdark\ as
decreasing with radius, that is only true in an absolute sense, just
like many other quantities decrease with radius (galactocentric
distance). Assuming that \kdark\ is not attributable to optically thick
\Hi, a roughly constant mass fraction of molecular material is CO dark,
independent of radius. This is in agreement with the findings of
\citet{Wolfire.2010} where they model the dark gas as the region
surrounding molecular clouds where the CO is photo-dissociated but not
the \Ht. This is in excellent agreement with our observations.

It is worth noting that there is no reason to think that the amount of
gas not traced by CO or \Hi\ should be constant.
Figure~\ref{fig.NHiDust} shows the dust surface density as a function of
the \Hi\ column density for 3 macropixels near the center and 3
macropixels between 4 and 5 kpc from the center. Examining the central
pixels, it is immediately apparent that the intercept (\kdark), varies
significantly from one pixel to another, even for neighboring regions.
Comparing with the lower panel, we see that \kdark\ tends to be lower in
the outskirts although for example, for the brown dots the distribution
is rather flat (moderate \kdark, infinite \gdr) at least when only the
\Hi\ is plotted. Assuming no CO is present at low \Hi\ column density,
it is also immediately apparent that there is more dust per unit gas
near the center, which is the equivalent of a radially increasing \gdr.
The low (high) \gdr\ is a factor common to all three pixels at small
(large) radii.
\FigNHiDust{}

\section{Conclusions}

In order to investigate how \gdr, \xco, and \kdark\ vary in M33, the
first step was to take a published estimate of the gas column density
$\NHtot_{dust}$ based on the Herschel dust observations and plot
$\NHtot_{dust}-\NHi$ versus \ico. The systematically positive intercept
(Fig.~\ref{fig.nht_ico_nocut_sub}) suggests that there is low-column
density gas traced by dust but not CO or \Hi, which we refer to as
$\kdark$ \citep{Tielens.1985, Planck-Collaboration.2011}.

The next step is to construct a map of the dust surface density. Two
methods were used -- the classical $\beta=2$ dust emissivity
(Fig.~\ref{fig.dust}, left panel) and the variable-$\beta$ (same Fig.,
right panel) developed by \citet{Tabatabaei.2014}. We adopt the second
method because in other subsolar metallicity galaxies
\citep{Galliano.2011} the classical approach yields too large a dust
mass, presumably due to a change in grain properties with respect to
Milky Way dust. Using $\beta=2$ for M33 also yields a very high dust
mass and \citet{Tabatabaei.2014} show that $\beta=2$ is a poor
approximation for M33.

We then look for optimal values of \gdr, \xco, and \kdark\ to relate the
dust surface density to the \Hi\ and CO intensities. Except where the
signal-to-noise ratio is high, major degeneracies are present between
these parameters (Fig.~\ref{fig.scatter_mean_sub}) such that they all
increase (or decrease) simultaneously with similar scatter in
$\log(\gdr)$.

Using simulated data with noise, a similar effect is seen in that the
deduced solutions generally have lower \gdr, \xco, and $\kdark$ than the
input values (Fig.~\ref{fig.LSCenter}--\ref{fig.LSOuterCut}).
Setting \gdr\ to the correct (input) value yields reasonably accurate
results. Solving only for \gdr\ and \xco, implicitly assuming $\kdark =
0$ when the input value was $\kdark = 5$ \Msunpsqpc, yields results for
\gdr\ and \xco\ that strongly depend on the amount of CO with respect to
\Hi. The degeneracies are illustrated by Figs~\ref{fig.LSnoise} and
\ref{fig.LSimu}.

An extremely computation-intensive simulation using the Bayesian
errors-in-variables approach was used to obtain ``true'' values of the
parameters. Fortunately, a very similar result can be obtained using the
Bayesian formalism but without the errors-in-variables approach, as
shown from the comparison in Fig~\ref{fig.compare}. The main difference
is the slightly lower uncertainty with the errors-in-variables approach.
The degeneracies present using the other methods are (almost) no longer
an issue (Fig.~\ref{fig.Correlnew}).
\FigCorrelNew{} 

There is a radial increase in \gdr\ from $\sim 200$ near the center to
nearly 400 in the outer disk. The \xco\ ratio remains constant with
galactocentric distance, as does the CO(2--1)/CO(1--0) line ratio
\citep{Druard.2014} and CO(3--2)/CO(2--1) line ratio (in prep.), unlike
what is observed in large spirals. The surface density of dark gas,
$\kdark$, decreases from the center (10\Msunpsqpc) to the outer parts
(roughly zero) in the same way as the CO emission such that the dark gas
represents close to half of the \Ht\ assuming that the dark gas is in
fact \Ht. As a result, the ratio of all \Ht\ (dark gas plus the \Ht\
traced directly by CO), is about twice the local value of $2 \times
10^{20}\Xunit$.

Some traces of the degeneracies between $\kdark$ and \gdr\ are still
present in that some macropixels with little CO find optimal values that
are physically unrealistic (typically \gdr\ $\sim$ 5000 with a
corresponding divergence of $\kdark$). Limiting the \gdr\ to values less
than 500 (5 times the Milky Way value) avoids the problem.

Overall, our results argue for a fairly high \gdr\ in M33 (\gdr\ $\ge
200$), a radially decreasing $\kdark$ roughly proportional to the amount
of CO emission, and a fairly constant \xco\ conversion both of the \Ht\
directly traced by CO and the total \Ht\ content including the dark gas
(whose radial distribution is similar to that of the CO).

The results presented here on the link between CO and total molecular
gas mass (and/or any optically thick \Hi) confirm the earlier estimates
of the \Ht\ mass of M33. As a result, either the \Ht\ is converted into
stars more quickly than in large spirals or the star-formation rate is
overestimated due to for example a change in IMF in this environment.
\begin{acknowledgements}
	PG thanks ERC starting grant (3DICE, grant
	agreement 336474) for funding during this work. PG's current
	postdoctoral position is funded by INSU/CNRS.
\end{acknowledgements}

\bibliographystyle{aa}
\bibliography{/Users/gratier/Documents/Biblio/biblio}

\end{document}